\documentclass[10pt,aps,prd,nofootinbib,superscriptaddress]{revtex4}
\usepackage{amssymb,amsmath,latexsym,graphics, graphicx,epsfig} 
\usepackage{latexsym}
\usepackage{amssymb}
\usepackage{epsfig,amsmath,graphics}
\usepackage{multirow}
\usepackage{appendix}
\usepackage{hyperref}
\usepackage{slashed}
\usepackage[boxsize=2em,aligntableaux=center]{ytableau}
\usepackage{dsfont}

\newcommand{\gsim}{\lower.7ex\hbox{$\;\stackrel{\textstyle>}{\sim}\;$}}
\newcommand{\lsim}{\lower.7ex\hbox{$\;\stackrel{\textstyle<}{\sim}\;$}}

\def\HH{{\cal H}}
\def\LL{{\mathcal L}}
\def\OO{{\cal O}}

\newcommand{\Tr}{{\text{ Tr }}}

\newcommand{\bef}{\begin{figure}[htbp]\begin{center}}
\newcommand{\eef}{\end{center}\end{figure}}

\newcommand{\bea}{\begin{eqnarray}}
\newcommand{\eea}{\end{eqnarray}}

\newcommand{\yss}{\ytableausetup{boxsize=0.5em}}
\newcommand{\ysn}{\ytableausetup{boxsize=2em}}

\newcommand{\fund}{\yss\ydiagram{1}\ysn}

\begin{document}

\pagestyle{plain}

\title{TASI Lectures on the Strong CP Problem and Axions}
\author{Anson Hook}

\begin{abstract} % abstract
These TASI lectures where given in 2018 and provide an introduction to the Strong CP problem and the axion.  I start by introducing the Strong CP problem from both a classical and a quantum mechanical perspective, calculating the neutron eDM and discussing the $\theta$ vacua.  Next, I review the various solutions and discuss the active areas of axion model building.  Finally, I summarize various experiments proposed to look for these solutions.
\end{abstract}
 \maketitle

\tableofcontents

\section{Introduction}

In these lectures, I will review the Strong CP problem, its solutions, active areas of research and the various experiments searching for the axion.
The hope is that this set of lectures will provide a beginning graduate student with all of the requisite background need to start a Strong CP related project.  As with any introduction to a topic, these notes will have much of my own personal bias on the subject so it is highly encouraged for readers to develop their own opinions.  I will attempt to provide as many references as I can, but given my own laziness there will be references that I miss.  As a result an exhaustive literature search is left as an exercise for the reader.

The Strong CP problem is in some sense both one of the most and least robust problems of the standard model (SM).
Unlike the flavor problem, but like the Higgs mass hierarchy problem, the Strong CP problem involves a parameter $\overline \theta$, 
which when sent to zero does not have an enhanced symmetry.  Thus it natural to expect it to have an $\OO(1)$ 
value.  It is sometimes said that the Strong CP problem is even more robust than the hierarchy problem because it is the 
only puzzle of the SM for which there is no anthropic solution.  Thus even people with the most extreme 
position on what can constitute a problem need to have an opinion of some sort on the Strong CP problem.

In some sense it is also one of the least robust problems of the standard model, because if $\overline \theta$ is set to be small at some scale, then it stays small by renormalization group (RG) evolution.  In some sense, one can just set it to be small and forget about it.  However, this property is unique to the minimal SM and doesn't hold in most of its extensions.  In the MSSM, $\overline \theta$ has 1 loop RG running from the gluino mass phase and this ``set it and forget it" approach fails.

Partly due to the robust nature of the Strong CP problem, solutions to it have always been of interest.  There are several standard symmetry-based solutions to the problem as well as the axion solution.  While none of these
have particularly convincing UV completions, some of the effective field theories (EFTs) are very economical and simple.  Most model building these days focuses on variations of the axion and its dark matter aspects.

Finally, a hot new topic is designing experiments to look for the axion or its variants.  Each experiment has different sensitivities to different regions of parameter space.  As there exist good reviews with excruciating amounts of detail about past, current and future experiments, what I will endeavor to do in these notes is to explain the basic idea behind each experiment.

In Sec.~\ref{Sec: classical}, I present a classical description of the Strong CP problem and some of its solutions.
In Sec.~\ref{Sec: quantum}, I present the Strong CP problem at the quantum level.
I then discuss the vacuum structure of QCD and how it resolves various historical confusions in Sec.~\ref{Sec: thetavac}.
I discuss non-axion solutions to the Strong CP problem in Sec.~\ref{Sec: not axion} and axionic solutions in Sec.~\ref{Sec: axion}.
The various dark matter aspects of axions are discussed in Sec.~\ref{Sec: dark matter}.
Finally I give a theorist's overview of experiments in Sec.~\ref{Sec: experiments}.

\section{The Strong CP problem and its solutions at the classical level} \label{Sec: classical}

\subsection{The Strong CP problem}

\begin{figure}
  \centering
  \includegraphics[width=7.5cm]{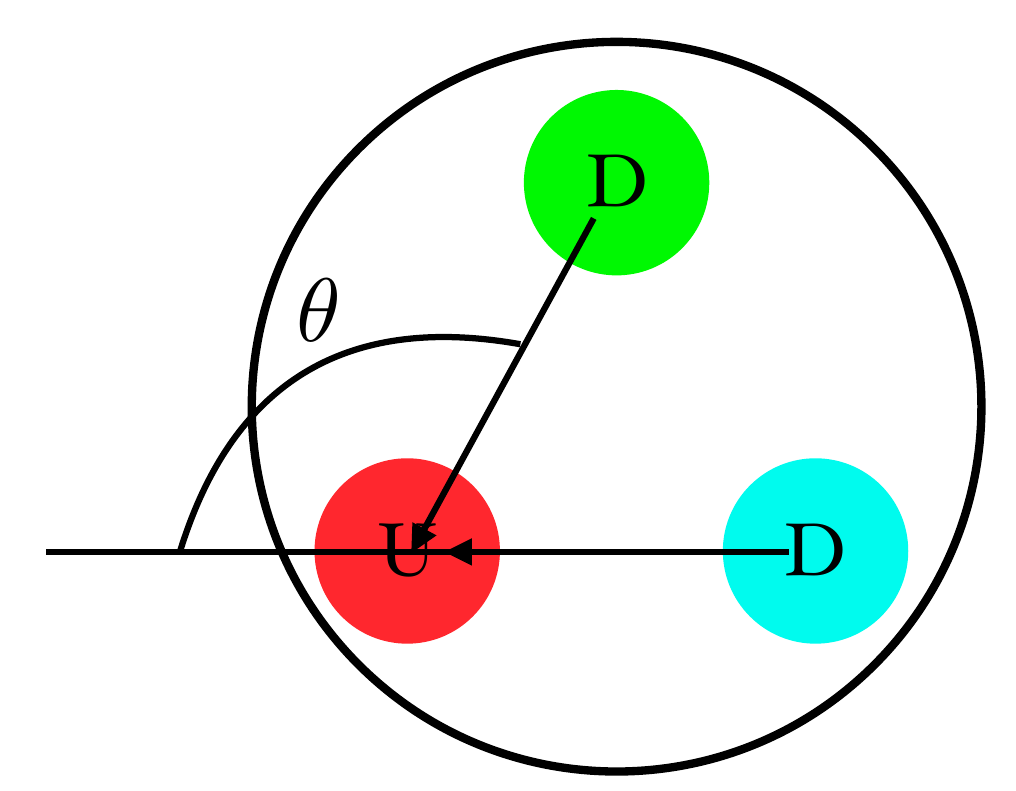}
  \caption{
    A classical picture of the neutron.  From this picture, an estimate of the neutron eDM may be made.}
    \label{Fig: Neutron}
\end{figure}

At its heart, the Strong CP problem is a question of why the neutron electric dipole moment (eDM) is so small.  It turns out that both the
problem and all of the common solutions can be described at the classical level.  Classically, the neutron can be thought of
as composed of a single charge $2/3$ up quark and two charge $-1/3$ down quarks.  Asking a student to draw the neutron usually ends up with something similar to that in Fig.~\ref{Fig: Neutron}.  If asked to calculate the eDM of the neutron, the student would simply take
the classical formula
\bea
\vec d = \sum q \vec r .
\eea
Using the fact that the neutron has a size $r_n \sim 1/m_\pi$, the student would then arrive at the classical estimate that
\bea
| d_n| \approx 10^{-13} \sqrt{1 - \cos \theta} \, e \, \text{cm}
\eea
Thus we have the natural expectation that the neutron eDM should be of order $10^{-13} e$ cm.  Because
eDMs are a vector, they need to point in some direction.  The neutron has only a single vector which breaks
Lorentz symmetry, and that is its spin.  Thus the eDM will point in the same direction as the spin (possibly with a
minus sign).  

Many experiments have attempted to measure the neutron eDM and the simplest conceptual way to do so is via a precession experiment.  Imagine that an unspecified experimentalist has prepared a bunch of spin up neutrons all pointing in the same direction.  The experimentalist then applies a set of parallel electric and magnetic fields to the system, which causes Larmor precession at a rate of
\bea
\nu_\pm  = 2 | \mu B \pm d E |.
\eea
After some time $t$, the experimentalist turns off the electric and magnetic fields and measures how many of the neutrons have
precessed into the spin-down position.  This determines the precession frequency $\nu_+$.  The experimentalist then redoes the
experiment with anti-parallel electric and magnetic fields.  This new experiment determines the precession frequency $\nu_-$.
By taking the difference of these two frequencies, the neutron eDM can be bounded.  The current best measurement of the
neutron eDM is~\cite{Baker:2006ts,Afach:2015sja,Graner:2016ses}
\bea
| d_n | \leq 10^{-26} e \, \text{cm} .
\eea
We have thus arrived at the Strong CP problem, or why is the angle $\theta \leq 10^{-13}$?  Phrased another way,
the Strong CP problem is simply the statement that the student should have drawn all of the quarks
on the same line!

\subsection{Solutions}

There are three solutions to the Strong CP problem that can be described at the classical level.  The first requires that parity
be a good symmetry of nature.
Under parity, space goes to minus itself.
\bea
P : \quad \vec x \rightarrow -\vec x.
\eea
We first consider a neutron whose spin and eDM point in the same direction, $ \hat s = \hat d_n $.  Remembering that angular momentum is $\vec s = \vec r \times \vec p$, we have under parity,
\bea
P : \quad d \rightarrow -d ,  \qquad s \rightarrow s.
\eea
Thus a neutron is taken from $ \hat s = \hat d_n $ to $ \hat s = -\hat d_n $ under parity.  We have studied the neutron and it is
an experimental fact that there is only a single neutron whose spin is 1/2.  Thus the only option is for the neutron to 
go to itself under parity.  The only way for both $ \hat s = \hat d_n $ and $ \hat s = -\hat d_n $ to be true is if the dipole moment is zero.
This is the parity solution to the Strong CP problem.  However, experimentally we have observed that parity is maximally broken by the weak
interactions.  Thus it is a bad symmetry of nature and any application to the Strong CP problem is necessarily more complicated.

The second classical solution is time-reversal (T) symmetry, typically called charge parity (CP) symmetry due to the fact that the combined CPT symmetry is a good symmetry of nature.  Under time reversal,
\bea
T : \quad t \rightarrow -t.
\eea
Considering again a neutron whose spin and eDM point in the same direction, $ \hat s = \hat d_n $, we find that under time reversal,
\bea
T : \quad d \rightarrow d  , \qquad s \rightarrow -s.
\eea
As before, a neutron is taken from $ \hat s = \hat d_n $ to $ \hat s = -\hat d_n $.  By the same reasoning, this again means that the neutron eDM must be zero.
As with parity, CP or equivalently T is not a symmetry of nature and is in fact maximally broken since the CP-violating phase in the CKM matrix is about $\pi/3$.

The last solution that can be seen at the classical level is the axion solution.  The situation of having two negative charges on opposite sides of a
positive charge seems very natural, just look at $CO_2$.  The plus charged carbon is exactly between the two oxygens with the equilibrium condition
being that the angle between the two bonds is exactly $\pi$ or in terms of the angle $\theta =0$.  The critical idea for making this situation work is that the angle between the two bonds is
dynamical.  If the initial angle is not $\theta = 0$, it quickly relaxes to $0$.  Motivated by this example, the axion solution is the idea that the angle $\theta$ is dynamical and can change.  It can be proven that the minimum will always be at $\theta = 0$~\cite{Vafa:1984xg} and the Strong CP problem is solved.

\begin{figure}
  \centering
  \includegraphics[width=7.5cm]{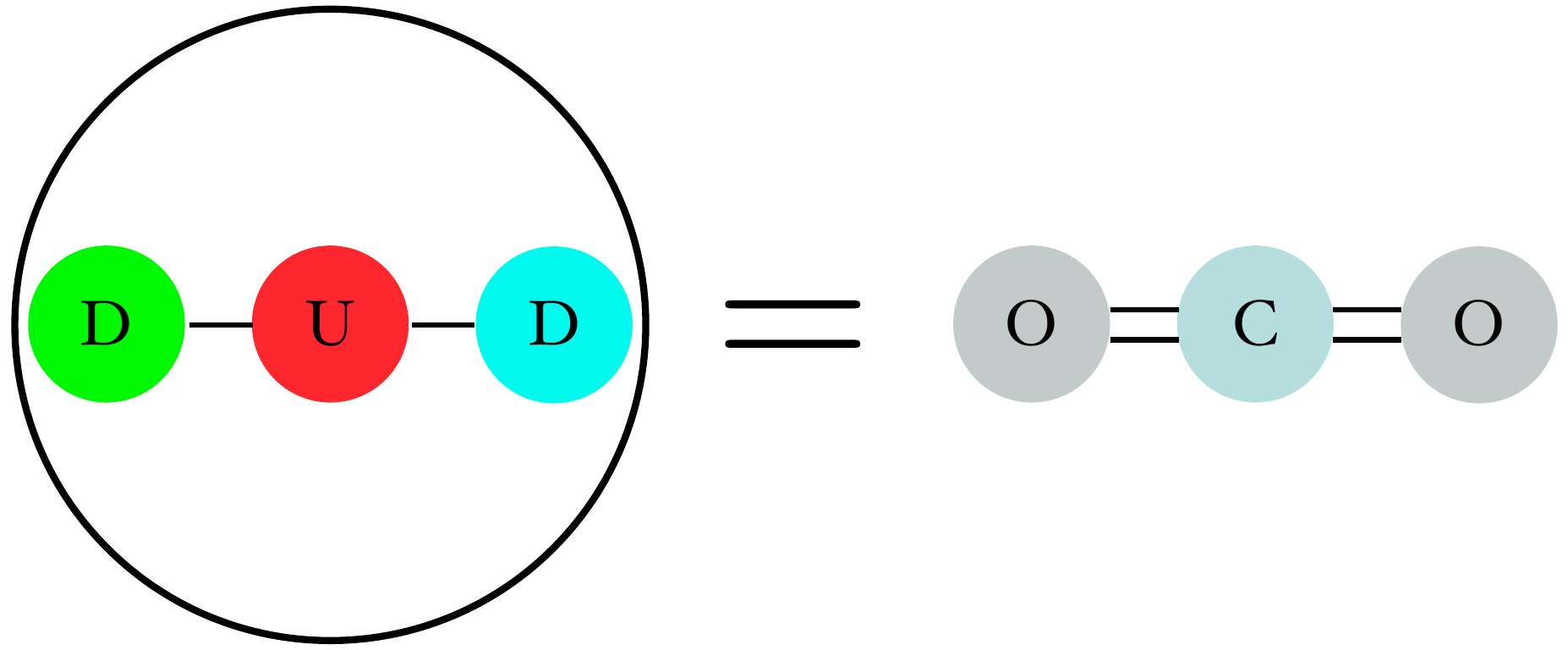}
  \caption{
    A axion solution to the Strong CP problem is treating the neutron like CO$_2$.  If the angle between the up and down quarks is dynamical, it will relax itself to the minimum energy configuration that has no dipole moment.  This dynamical angle is called the axion.}
    \label{Fig: Neutron}
\end{figure}

\section{The Strong CP problem at the quantum level} \label{Sec: quantum}

We now formulate the Strong CP problem at the quantum level.  As the Strong CP problem is a question about the properties of the neutron,
we need to develop a theory of neutrons and low-energy QCD.  In this section, we follow a semi-historical route.  We first describe how to get low-energy QCD, aka the theory of pions, incorrectly.  We then
fix it via a better understanding of anomalous symmetries.  Next, we describe how to get the theory of pions correctly.    Finally, we add neutrons into the
theory and calculate the neutron eDM.

\subsection{Low-energy QCD done incorrectly}

We consider QCD with two light flavors.  This theory has gluons ($A_\mu$), left-handed quarks ($q = (u \, d)$) and right-handed quarks ($q^c = (u^c \, d^c)$).  The fermions $q$ and $q^c$ are Weyl fermions.  For those unfamiliar or in search of a review of Weyl fermions, both Ref.~\cite{Martin:1997ns} and Ref.~\cite{Dreiner:2008tw} provide good introductions to the topic.
Aside from the kinetic terms, the theory has the Lagrangian
\bea
\mathcal{L} \supset \frac{\theta g_s^2}{32 \pi^2} G \tilde G + q M q^c , \qquad  M = \begin{pmatrix} 
  m_u & 0  \\
  0  & m_d 
 \end{pmatrix}
\eea
where $\tilde G^{\mu \nu} = \frac{1}{2} \epsilon^{\mu \nu \rho \sigma} G_{\rho \sigma}$.
$\theta$ plays no roll in this subsection and will be ignored for now.
This theory has an $SU(3)$ gauge group and 4 global symmetries $SU(2)_L \times SU(2)_R \times U(1)_B \times U(1)_A$.  Under these symmetries, the particles and spurions transform as
\begin{center}
\bea
\begin{tabular}{c|c|cccc}
&$SU(3)$&$SU(2)_L$&$SU(2)_R$&$U(1)_B$&$U(1)_A$\\
\hline
&&\\[-8pt]
$A_\mu$ & \text{adj}  &  & &  &  \\
$q$ & $\fund$ & $\fund$ & & 1 & 1 \\
$q^c$ & $\overline \fund$ &  & $\overline \fund$ & -1 & 1 \\
$M$ &  & $\overline \fund$ & $ \fund$ &  & -2 \\
\end{tabular}
\eea
\end{center}

At low energies, this theory becomes strongly coupled and we have no analytic traction on what happens.  Instead,
what we will do is use various inputs from experiment to build an effective field theory of the pions.  The starting point
is the measured fact that QCD confines.  In particular, it has been determined experimentally that 
\bea
\langle q q^c \rangle \ne 0 ,
\eea
which breaks $SU(2)_L \times SU(2)_R$ down to its diagonal $SU(2)_D$, and also breaks $U(1)_A$.  As with any spontaneous symmetry breaking,
there will exist Goldstone bosons: These are called the pions and are expressed in terms of the matrix
\bea
U = e^{i \frac{\Pi^a}{\sqrt{2} f_\pi} \sigma^a} ,
\eea
where $\sigma^{1-3}$ are the Pauli spin matrices and $\sigma^0$ is the identity matrix.  $\Pi^0$ is associated with the breaking of $U(1)_A$ and is called the $\eta'$ boson.  Meanwhile the other pions are typically referred to as $\Pi^3 = \pi^0$ and $\Pi^{1,2} = \pi^{1,2}$.  $U$ is a unitary matrix so that $U U^\dagger$ is the identity matrix.
$U$ has the symmetry transformation properties
\begin{center}
\bea
\begin{tabular}{c|cccc}
&$SU(2)_L$&$SU(2)_R$&$U(1)_B$&$U(1)_A$\\
\hline
&&\\[-8pt]
$U$ & $\fund$ & $\overline \fund$ &  & 2 \\
\end{tabular}
\eea
\end{center}
I'll leave it as an exercise to the reader to demonstrate that the vev of $U$ preserved the diagonal group $L = R$ while breaking the axial group $L = R^\dagger$.

As we know nothing of how we got to this theory, we will write down all renormalizable operators consistent with symmetries with arbitrary coefficients.
The leading order operator that one can write down is
\bea
\mathcal{L} = f_\pi^2 \Tr \partial_\mu U \partial^\mu U^\dagger = \frac{1}{2} \partial_\mu \pi^a \partial^\mu \pi^a + \cdots
\eea
All other terms in the potential are higher-dimensional operators and their coefficients are fixed by the requirement that when $U$ is expanded in terms of the pion fields, the kinetic term is canonically normalized.  We now include the mass of the quarks, keeping only the leading-order operator.  In other words, we perform a series expansion in small masses.  Remembering that the mass matrix has transformation properties under the flavor symmetries, we write the leading-order operator as
\bea
\mathcal{L} = f_\pi^2 \Tr \partial_\mu U \partial^\mu U^\dagger + a f_\pi^3 \Tr M U + h.c. ,
\eea
where a is an arbitrary constant that will be determined by matching with data.
Expanding this Lagrangian in terms of the pion fields, one obtains the mass matrix
\bea
V = a f_\pi (m_u + m_d) \pi^+ \pi^- + \frac{a f_\pi}{2} \begin{pmatrix}
  \pi^0  & \eta' 
 \end{pmatrix} \begin{pmatrix}
  m_u + m_d & m_u - m_d  \\
  m_u - m_d  & m_u + m_d 
 \end{pmatrix} \begin{pmatrix}
  \pi^0  \\
  \eta' 
 \end{pmatrix} , \qquad \pi^\pm = \frac{\pi^1 \pm i \pi^2}{\sqrt{2}} .
\eea
We see that there are four light particles whose masses obey $2 m_{\pi^+} = m_{\pi^0} + m_{\eta'}$.  At this point, we again turn to experiment and find that $m_{\pi^+} \approx m_{\pi^0} \approx 140$ MeV while $m_{\eta'} \approx 960$ MeV.  This clearly does not obey the sum rule that the EFT just derived, so something has gone wrong.  As we will discuss in the next section, it turns out that $U(1)_A$ is actually not a good symmetry and that the $\eta'$ boson obtains a large mass from another source.

\subsection{Anomalous symmetries}

In this subsection, we discuss how the $U(1)_A$ symmetry discussed above is actually not a good symmetry of the theory and the implications of it.  
From any number of QFT textbooks, e.g. Ref.~\cite{Peskin:1995ev}, one finds that if one rotates the quarks by
\bea
u \rightarrow e^{i \alpha} u , \qquad u^c \rightarrow e^{i \alpha} u^c ,
\eea
then under this rotation, the Lagrangian also changes as
\bea
\label{Eq: ano}
\LL \rightarrow \LL + \alpha \frac{g^2}{16 \pi^2} G \tilde G .
\eea
The reason for this anomalous symmetry is that the measure is not invariant under this transformation.

Because there is no symmetry, there should be no Goldstone boson.  However, explicitly broken symmetries are still useful.  After all, in the previous subsection, we showed how to start building a theory of pions even when there are explicit mass terms that break the symmetries.  The star of the previous show were spurions, constants that transform under symmetries.  Thus, we wish to find a constant under which we can take this non-symmetry and turn it into a spurious symmetry.  This particular example is usually called an anomalous symmetry due to the association with the anomaly in Eq.~\ref{Eq: ano}.

By remembering that there was a term in the Lagrangian that is
\bea
\LL \supset \theta \frac{g^2}{32 \pi^2} G \tilde G ,
\eea
we see that we can cancel the piece added to the Lagrangian in Eq.~\ref{Eq: ano} by shifting $\theta$.  $\theta$ is now our spurion.
For 2-flavor QCD, the proper anomalous symmetry is
\bea
u \rightarrow e^{i \alpha} u , \qquad d \rightarrow e^{i \alpha} d , \qquad \theta \rightarrow \theta - 2 \alpha .
\eea

Note that there are several important differences between $\theta$ as a spurion and $M$ as a spurion.  
A major difference is that $\theta$ realizes the symmetry non-linearly, i.e. it shifts under the symmetry rather than changing multiplicatively the way $U(1)_A$ acts on the pion matrix $U$.  To obtain a spurion that transforms linearly, we let $\theta$ appear in the Lagrangian as $e^{i \theta}$.

For the spurion $M$, the masses of the pseudo-Goldstone bosons are suppressed by $M$ in the
$M \rightarrow 0$ limit.  The reason is that the symmetry is restored in the $M \rightarrow 0$ limit so that the Goldstone boson masses must go to zero in this limit.  Thus we can take $M$ small and apply a Taylor series.  However, this sort of expansion is impossible for $\theta$ because $|e^{i \theta}|=1$, so even if $\theta = 0$, the pseudo-Goldstone boson mass is still non-zero.  This is reflected in the fact that $\theta =0$  does not convert the anomalous symmetry into a true symmetry.  The anomalous symmetry never was and never will be a symmetry of the theory~\footnote{If $g_s = 0$ then the anomalous symmetry would be a good symmetry, but then confinement does not occur.}.   Despite this, it still has its uses, as we will see in the next subsection.

\subsection{The theory of pions and neutrons done properly}

As mentioned in the previous subsection, the $U(1)_A$ symmetry is not a good symmetry of nature.  Recall that the anomalous symmetry is
\bea
u \rightarrow e^{i \alpha} u , \qquad d \rightarrow e^{i \alpha} d , \qquad \theta \rightarrow \theta - 2 \alpha .
\eea
Because a constant of nature, $\theta$, transforms under this symmetry, the corresponding pseudo-Goldstone boson, $\eta'$, obtains a mass
even in the limit where the quark masses go to zero.

As in the case of non-zero quark masses, broken symmetries are still useful in constraining how their corresponding pseudo-Goldstone boson appears.  To see how $\eta'$ transforms under $U(1)_A$, we note that
$q \rightarrow e^{i \alpha} q$ tells us how $U \propto q q^c$ transforms.  Thus there is an anomalous symmetry
\bea
U \rightarrow e^{i \alpha} U , \qquad \theta \rightarrow \theta - 2 \alpha , \qquad M \rightarrow e^{-i \alpha} M.
\eea
Written in terms of the $\eta'$ boson, this means that the following is a good symmetry of the theory :
\bea
\label{Eq: anomalous}
\eta' \rightarrow \eta' + \sqrt{2} \alpha f_{\eta'} , \qquad \theta \rightarrow \theta - 2 \alpha ,  \qquad M \rightarrow e^{-i \alpha} M .
\eea

Now armed with the fact that $U(1)_A$ is not a good symmetry of nature, we can write down a new term in the effective Lagrangian :
\bea
\label{Eq: etap}
\mathcal{L} = f_\pi^2 \Tr \partial_\mu U \partial^\mu U^\dagger + a f_\pi^3 \Tr M U + b f_\pi^4 \det U + h.c.
\eea
which is invariant under $SU(2)_L \times SU(2)_R \times U(1)_B$ but not invariant under $U(1)_A$.  But that is fine because $U(1)_A$ was
never a true symmetry to begin with.  Note that while $U(1)_A$ isn't a good symmetry, the anomalous symmetry given in Eq.~\ref{Eq: anomalous} is still valid.  Thus we see that the phase of the complex coefficient $b$ is fixed to be
\bea
b = |b| e^{i \theta} .
\eea
The mass of the $\eta'$ boson can be obtained by Taylor expanding Eq.~\ref{Eq: etap} as
\bea
\LL = \frac{1}{2} m_{\eta'}^2 \left( \eta' - \frac{\theta f_{\eta'}}{\sqrt{2}} \right)^2 + \cdots
\eea
Plugging this expectation value into the matrix $U$, we find that
\bea
U = e^{i \frac{\theta}{2}} e^{i \frac{\pi^a}{\sqrt{2} f_\pi} \sigma^a} .
\eea

Now that we understand how the $\eta'$ behaves, we can finally go back and redo the theory of pions carefully.
The first step is to find the vacuum about which to expand.  This vacuum can be non-trivial.  The easiest way to see this
is to look at the pion masses, $m_\pi \sim m_u + m_d$.  If the quark masses were negative, then the pion mass would
also be negative.  To find the vacuum state, we assume that $\pi^0$ has an expectation value $\langle \pi^0 \rangle = \phi \sqrt{2} f_\pi$.
It is left as an exercise to the reader to show that the charged pions do not obtain an expectation value.  Thus we are expanding about
\bea
U = \begin{pmatrix}
  e^{i \phi + i \theta} & 0  \\
  0  & e^{- i \phi + i \theta}
 \end{pmatrix} .
\eea
$\phi$ comes from the expectation value of $\pi^0$, while $\theta$ appears due to the expectation value of $\eta'$.

The potential comes from the term
\bea
V = - a f_\pi^3 \Tr \left ( \begin{pmatrix}
  m_u e^{i \theta_u} & 0  \\
  0  & m_d e^{i \theta_d}
 \end{pmatrix} U  \right ) + h.c. = - 2 a f_\pi^3 \left [ m_u \cos ( \phi + \frac{\overline \theta}{2} ) + m_d \cos ( \phi - \frac{\overline \theta}{2} ) \right ] ,
 \eea
 \bea
 \label{Eq: theta}  \overline \theta = \theta + \theta_u + \theta_d ,
\eea
where we have used a shift of $\phi$ to express the potential in a clean form.  Note that $a$ is necessarily real because the QCD Lagrangian is CP conserving up to the mass terms of the quarks (i.e. $\theta_u$ and $\theta_d$ can be non-zero) and $\theta$.  Thus all CP-violating effects in QCD itself come from the $\theta$ term,
and how $\theta$ appears is restricted by the anomalous $U(1)_A$ symmetry.  The mass term is $U(1)_A$ invariant so that there is no $\theta$ dependence and thus the arbitrary constant $a$ must be real.  As the reader can check, whether $a$ is positive or negative has no physical effect, so for simplicity we take it to be positive.
On the other hand, the masses can break CP with their non-trivial phases so we have written them out explicitly.

With a little bit of effort, the minimum of this potential can be found to be
\bea
\label{Eq: thetapot}
\tan \phi = \frac{m_u - m_d }{m_u + m_d} \tan \frac{\overline \theta}{2} , \qquad V = - m_\pi^2 f_\pi^2 \sqrt{1 - \frac{4 m_u m_d}{(m_u + m_d)^2} \sin^2 \frac{\overline \theta}{2}} .
\eea
Expanding about this minimum, we find that the pion masses are
\bea
m_{\pi}^2 = a f_\pi \sqrt{m_u^2 + m_d^2 + 2 m_u m_d \cos \overline \theta}.
\eea
Amusingly, notice that the leading order contribution to the pion mass vanishes when $m_u = m_d$ and $\theta = \pi$.  It is a fun exercise to show that this is the result of isospin.
%While at it, the reader can also repeat this calculation for the 3 quark case and show how the sum rules change as a function of $\theta$ and argue that $\theta$ must be small from
While the reader is at it, I also encourage them to also attribute the mass difference between the charged and neutral pions to the quadratic divergence due to the electric charge of the $\pi^+$.  The particle that cuts off this divergence is the
$\rho$ meson.  The standard quadratic divergence estimate for the mass difference cut off by the $\rho$ meson should reproduce the measured difference in mass between the $\pi^+$ and $\pi^0$, lending credence to standard arguments for quadratic divergences.

After this long and arduous trek, we finally have a theory of pions that gives the correct pion masses.  We can now
incorporate protons and neutrons into the theory.  Again appealing to experiment, we know that protons and neutrons are
each composed of three quarks.  We can thus construct a nucleon field $N$.
\bea
N = q q q  = \begin{pmatrix}
  p  \\
  n
 \end{pmatrix} ,
\eea
with $N^c = q^c q^c q^c$.
I have not written down how the indices are contracted.  I leave as a fun exercise to the reader to contract the indices and show that the proton is made of two up quarks and a down quark, and that the proton and neutron fall into a $SU(2)$ doublet.

Working through the indices, $N$ ($N^c$) transforms as a doublet under $SU(2)_L$ ($SU(2)_R$).
As before, we now write down all of the leading-order terms with arbitrary coefficients :
\bea
\LL = - m_N N U^\dagger N^c - c_1 N M N^c - c_2 N U^\dagger M^\dagger U^\dagger N^c - \frac{i}{2} (g_A -1) \left [ N^\dagger \sigma^\mu U \partial_\mu U^\dagger N + N^{c,\dagger} \sigma^\mu U^\dagger \partial_\mu U N^c    \right ] .
\eea
Expanding these terms out to leading order in pions and integrating by parts, we find that the leading CP-preserving and violating interactions are
\bea
\LL = - \overline \theta \frac{c_+ \mu}{f_\pi} \pi^a N \tau^a N^c - i \frac{g_A m_N}{f_\pi} \pi^a N \tau^a N^c , \qquad \mu = \frac{m_u m_d}{m_u + m_d} .
\eea
Note that in Weyl notation, the difference between CP preserving and violating is whether the coupling is imaginary or real, and not the $\gamma^5$ matrices seen in Dirac notation.  $c_+ = c_1 + c_2$ gives a mass splitting between various nucleons and can be determined to be $c_+ \approx 1.7$ using the measured value of their masses.  $g_A$ gives the leading-order interaction between protons and neutrons, so by scattering protons off neutrons, we can measure $g_A \approx 1.27$~\footnote{Actually, $g_A$ is better related to the decay of the neutron, but that would be a long digression all by itself.}.

\begin{figure}
  \centering
  \includegraphics[width=7.5cm]{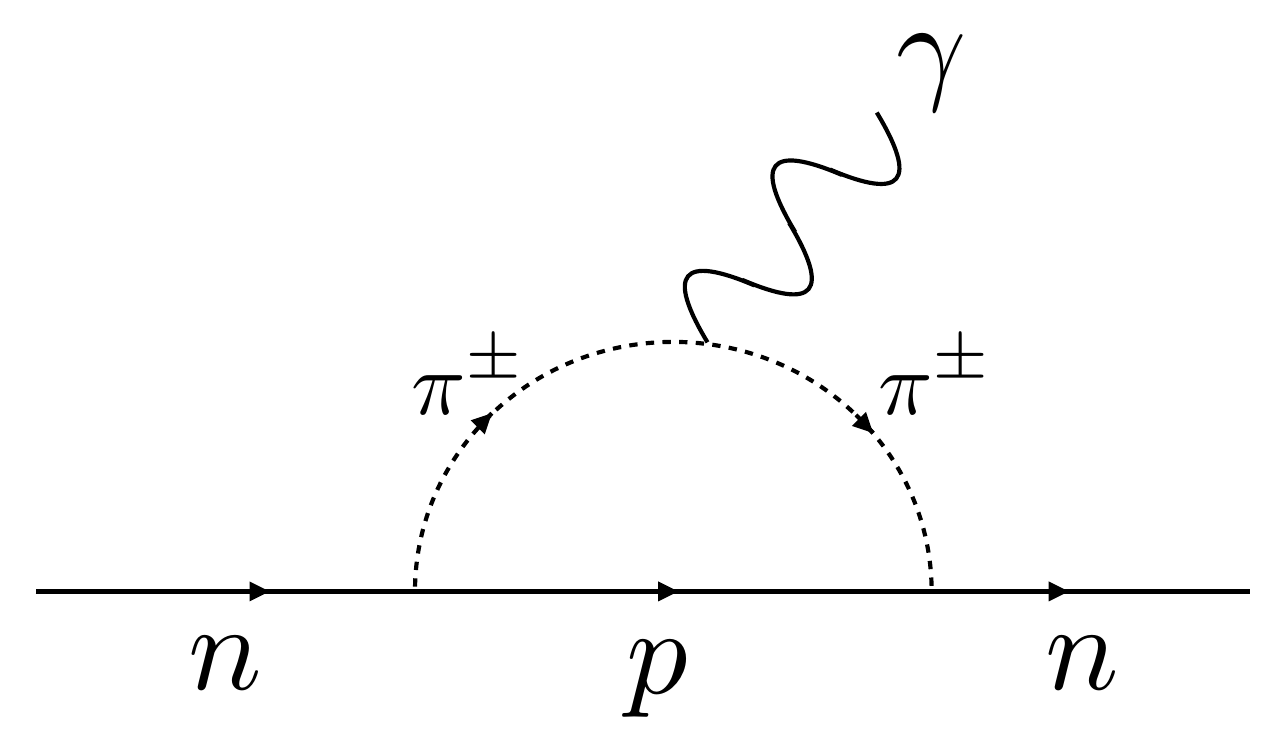}
  \caption{
    The Feynman diagram giving the leading-order contribution to the neutron eDM.}
    \label{Fig: feyn}
\end{figure}

To obtain the neutron eDM, we calculate the Feynman diagram shown in Fig.~\ref{Fig: feyn}.  There is not much to be learned from the computation itself, so I will only briefly sketch the procedure using Dirac notation.  For those interested in more details, see Ref.~\cite{Srednicki:2007qs}.
The matrix element of the Feynman diagram is
\bea
\nonumber
i M &\approx& -i e \frac{\sqrt{2} g_A m_N}{f_\pi} \frac{\sqrt{2} \overline \theta c_+ \mu}{f_\pi} \epsilon^*_\mu(q) \int \frac{d^4 l}{(2 \pi)^4} 2 l^\mu  \overline u(p') \frac{\left ( (-\slashed{l} - \slashed{p}/2 - \slashed{p}'/2  + m_N) \gamma_5 + \gamma_5 (-\slashed{l} - \slashed{p}/2 - \slashed{p}'/2  + m_N)   \right)}{((l+p/2+p'/2)^2 + m_N^2) ((l + q/2)^2 + m_\pi^2) ( (l - q/2)^2 + m_\pi^2)} u(p) \\
&\approx& \frac{e \overline \theta g_A c_+ \mu \log \frac{\Lambda^2}{m_\pi^2}}{4 \pi^2 f_\pi^2} \epsilon^*_\mu(q) \overline u(p') \gamma^{\mu \nu} q_\nu i \gamma_5 u(p) \label{Eq: edm} ,
\eea
where $p$ ($p'$) is the incoming (outgoing) momentum of the neutron, and q is the incoming momentum of the photon.  Anticipating that $\overline \theta$ is small, we have performed a Taylor series in $\overline \theta$ as well as taken the leading-order piece in $q$.  $\Lambda \sim 4 \pi f_\pi$ is the UV cutoff of our theory of pions.

Let us now pretend that the neutron has an eDM in the Lagrangian,
\bea
\LL \supset d_n F_{\mu \nu} \overline n \gamma^{\mu \nu} i \gamma_5 n .
\eea
This would correspond to a diagram with the matrix element
\bea
i M = 2 d_n  \epsilon^*_\mu(q) \overline u(p') \gamma^{\mu \nu} q_\nu i \gamma_5 u(p) .
\eea
Comparing this with Eq.~\ref{Eq: edm}, we see that
\bea
\label{Eq: neutronedm}
d_n = \frac{e \overline \theta g_A c_+ \mu}{8 \pi^2 f_\pi^2} \log \frac{\Lambda^2}{m_\pi^2} \sim 3 \times 10^{-16} \, \overline \theta \, \text{e cm} .
\eea
Finally, comparing $d_n$ to the bounds on the neutron eDM gives
\bea
\overline \theta \lesssim 10^{-10}
\eea

To show that these results are correct, we can perform an important check that the potential depends exclusively on $\overline \theta$ as defined in Eq.~\ref{Eq: theta}.
To see that $\overline \theta$ is the only physical quantity, we remind the reader of the QCD Lagrangian
\bea
\LL \supset m_u e^{i \theta_u} u u^c + m_d e^{i \theta_d} d d^c + \frac{\theta g_s^2}{32 \pi^2} G \tilde G .
\eea
Two anomalous symmetries constrain the theory :
\bea
u \rightarrow e^{i \alpha} u , \qquad \theta_u \rightarrow \theta_u - \alpha , \qquad \theta \rightarrow \theta + \alpha
\eea
and
\bea
d \rightarrow e^{i \alpha} d , \qquad \theta_d \rightarrow \theta_d - \alpha , \qquad \theta \rightarrow \theta + \alpha.
\eea
These anomalous symmetries are simply a reflection of how you're defining your quarks, so any
physical quantity is invariant under these anomalous symmetries.  It is easy to see that the only invariant quantity is $\overline \theta$ and thus any physical answer can only depend on $\overline \theta$.

Unfortunately, the Strong CP literature is often not clear about $\overline \theta$ versus $\theta$.  People (this author included) will often be sloppy in their notation and simply write $\theta$ when they mean $\overline \theta$.  While I will try to be careful in this review, the reader should be alert in general and use context to determine if the author means $\overline \theta$ or just $\theta$.

\section{The $\theta$ vacua} \label{Sec: thetavac}

In this section, we are motivated by two confusing puzzles whose solutions lie in what is known as the $\theta$ vacua.
The first is the following statement :  If we start with the Lagrangian
\bea
\label{Eq: theta1}
\LL \supset -\frac{1}{4} G^2 +  \frac{\theta g_s^2}{32 \pi^2} G \tilde G
\eea
and calculate the Hamiltonian $\HH$, we find that it is independent of $\theta$!  We already know that all of physics can be boiled down to
asking how a given state evolves in time, and that the answer is always given by
\bea
e^{i H t} \mid x_i \rangle .
\eea
Thus, all we need to know about a system are its initial conditions and the Hamiltonian.  If the Hamiltonian does not
depend at all on $\theta$, then $\theta$ should not affect any physical quantities.  This is the first puzzle.

The second puzzle is related to the first and stems from the fact that the coupling in Eq.~\ref{Eq: theta1} is a total derivative :
\bea
G \tilde G = \partial_\mu K^\mu , \qquad K^\mu = \epsilon^{\mu \nu \rho \sigma} A_\nu^a \left [ F^a_{\rho \sigma} - \frac{g}{3} f^{abc} A^b_\rho A^c_\sigma \right ] .
\eea
An alternative way to calculate physical quantities is to use the principle of least action,
\bea
\langle x_f \mid e^{i H t} \mid x_i \rangle = \int_{x_i}^{x_f} d[x] e^{i S} .
\eea
What is of interest is the action, which is the integral over all space of the Lagrangian.  Using the divergence theorem,
\bea
S = \int d^4 x \LL \supset \int d^4 x \frac{\theta g_s^2}{32 \pi^2} G \tilde G = \int d^3 x \frac{\theta g_s^2}{32 \pi^2} K^{\hat r}  \mid_{r \rightarrow \infty} .
\eea
Thus, if $K$ vanishes faster than $1/r^3$ at infinity, then this quantity will integrate to zero and $\theta$ cannot have any physical effect.
An expectation for how $K$ scales can be obtained using the fact that we are dealing with a system at finite energy :
\bea
\text{Energy} \sim \int d^3 x \left ( E^2 + B^2 \right ) < \infty , \qquad E < \frac{1}{r^2} ,
\eea
where E decreases faster than $1/r^2$ to have finite energy.  Via handwaving logic,
\bea
\partial K \sim K/r \sim G \tilde G \sim G^2 < \frac{1}{r^4} .
\eea 
Thus we expect $K$ to decrease faster than $1/r^3$, and the action to be independent of $\theta$.  This is the second puzzle.

The resolution of these puzzles lies in the instantons and the vacuum structure of QCD.  In particular, the first puzzle is solved by
the fact that $\theta$ appears in super selection rules in the Hamiltonian formalism and thus plays a role in the specification of the 
initial states.  The second puzzle is solved by the presence of finite-energy field configurations known as instantons where $K \sim 1/r^3$ and the surface integral does not vanish at infinity.
Instantons are a gnarly subject and require their own entire review (the classic reference is Ref.~\cite{Coleman:1985rnk} but there exist other more recent references such as Ref.~\cite{Forkel:2000sq}), so I will only sketch a brief outline.

Our starting point is asking what sort of boundary conditions to impose on our theory.  We want a system with finite energy,
so as $r \rightarrow \infty$, we need the gauge field to become pure gauge so that the $E$ and $B$ fields vanish :
\bea
r \rightarrow \infty , \qquad A_\mu \rightarrow U \partial_\mu U^\dagger.
\eea
Then we ask whether the gauge configurations that $A_\mu$ goes to at infinity are all equivalent, or to phrase it in the parlance of a mathematician, whether the mappings between the gauge group, $SU(3)$, and the sphere at infinity, $S_3$, are all equivalent~\footnote{For the more mathematically minded, this is a question of homotopy.}.  In particular, one can either prove or look up that $\pi_3(SU(3)) = \mathbb{Z}$.  The 3 in the subscript indicates mapping $SU(3)$ to $S_3$.  Thus, we discover that the gauge configurations that $A_\mu$ can go to at spatial infinity are characterized by an integer.  They cannot be smoothly deformed into each other without leaving pure gauge, as integers are not smoothly connected.

For a better understanding of this result, we consider the simpler example of $\pi_1(U(1))$, or how to map a circle to another circle.
Imagine that the first circle has an angle $\phi_1$ and the second circle has an angle $\phi_2$.  Being angles, they are $2 \pi$ shift symmetric.
When mapping one circle to the other circle, we have
\bea
\phi_1 \rightarrow n \phi_2 ,
\eea
where for simplicity we have mapped $\phi_1 = 0$ to $\phi_2 = 0$.  We see that this mapping is characterized by the integer $n$, typically called the winding number, which indicates how many times the first circle winds around the second.
$\pi_3(SU(3)) = \mathbb{Z}$ works in much the same way.  As $SU(3)$ contains the group $SU(2)$, which is the symmetry of a $S_3$ sphere, we see that mapping $SU(2)$ to $S_3$ is just the higher dimensional version of the previous example, which is again characterized by the winding number.

We have now discovered that the asymptotic behavior of the gauge field is characterized by an integer.  It turns out that the winding number of any given gauge configuration can be determined by
\bea
\label{Eq: winding}
\int d^4 x \frac{1}{32 \pi^2} G \tilde G = n_1 - n_2 .
\eea
Where $n_1$ is the winding number of the gauge field at infinity and $n_2$ is the winding number at the origin.  Thus, by specifying what $G \tilde G$ integrates to, we can dictate the asymptotic behavior of the gauge field configurations.  Eq.~\ref{Eq: winding} is a very important fact and more details can be found in Ref.~\cite{Forkel:2000sq}.

Since the vacuum is an eigenstate of energy which has the lowest energy, let's check if the field configurations specified by $n$ are
energy eigenstates.  The first thing to note is that all $n$ states have the same energy.  Each of these $n$ states are specified by how the gauge field falls off at infinity.  The lowest energy state in each of these sectors is when the gauge field takes on its asymptotic value at all points in space time.  Because the gauge field is pure gauge, they all
have zero E and B fields and thus the same energy.  Next we ask whether these states mix with one another under time evolution.  To answer that question, we first remind ourselves that physical quantities are obtained from
\bea
\int_{A_i}^{A_f} d[A] e^{i S} .
\eea
Imagine that our initial state is $n=0$ and we are looking for tunneling into the $n=1$ state.  Since our initial state has $n=0$, we can say that our initial state is the $A=0$ state.  In the path integral, we integrate over all finite-energy gauge-field configurations that can be reached by continuously deforming the $A=0$ initial state.  Rather than deriving the answer, we shall take a guess-and-check approach.  We simply guess the following gauge-field configuration, which is called an instanton :
\bea
A_\mu = \frac{r^2}{r^2 + \rho^2} g \partial_\mu g^{-1} , \qquad g = \frac{x_4 \mathbf{1} + i \vec x \cdot \vec \tau}{r} , 
\eea
where $\rho$ is an arbitrary constant that is the size of the instanton, $\tau$ are the Pauli spin matrices, and $g$ is pure gauge.
It is simple, but annoying, to check that this gauge-field configuration has finite energy and that
\bea
\int d^4 x \frac{1}{32 \pi^2} G \tilde G = 1 .
\eea
Hence this field configuration has all of the properties we need to show that there is non-zero overlap between the two states $n=0$ and $n=1$.  In fact, we can connect any two different $n$ states by superimposing these gauge-field configurations.

Now we will find the eigenstates of the Hamiltonian.  To get a simple intuition for the answer, we first pretend that there are a finite number
of states, $D$, mixing with each other.  The matrix that we are diagonalizing is called a circulant matrix and is of the form
\bea
\begin{pmatrix}
    1  & 2 & 3 & \cdots & D-1 
 \end{pmatrix}  \cdot 
 \begin{pmatrix}
  E & \epsilon_1 & \epsilon_2 & \cdots & \epsilon_{D-1}  \\
  \epsilon_{D-1}  & E & \epsilon_1 & \epsilon_2 & \cdots  \\
  . & . & . & . & .   \\  
  \epsilon_1  & & \epsilon_2 & \cdots & \epsilon_{D-1} & E 
 \end{pmatrix} \cdot \begin{pmatrix}
    1  \\ 2 \\ 3 \\ \cdots \\ D-1 
 \end{pmatrix} .
\eea
Its eigenvectors are completely independent of the values of $\epsilon$ and $E$, though its eigenvalues do depend on them.  The eigenvectors are
\bea
\begin{pmatrix}
    1  & w_j & w_j^2 & \cdots & w_j^{D-1} 
 \end{pmatrix}  , \qquad w_j = e^{2 \pi i j/D} .
\eea
They are simply the sum of each state weighted by a multiple of the $D$th root of unity.  We can now see that in the $D \rightarrow \infty$ limit, we  weight each of the states by a phase and sum.
The eigenstates of the Hamiltonian are
\bea
\mid \theta \rangle = N \sum e^{i \theta n} \mid n \rangle ,
\eea
where $N$ is just some normalization factor that doesn't matter.

$\theta$ is what is called a super-selection rule.  Because it is impossible to transition from one value of $\theta$ to another, 
we choose a single value of $\theta$ and throw out all other values when we define our theory.  Our sector, with it's value of $\theta$ is completely orthogonal to these other states, so there is no reason to include them in our theory.

We have seen that $\theta$ appears as part of the state in the Hamiltonian formalism.  Now we show that when we transition to path integral
formulation, $\theta$ also appears in the Lagrangian.  We note that
\bea
\langle \theta \mid \OO  \mid \theta \rangle &=& \sum_{m,n} e^{i \theta (m-n)} \langle m \mid \OO  \mid n \rangle = \sum_{\Delta,n} e^{i \theta \Delta} \langle n+\Delta \mid \OO  \mid n \rangle = \sum_{\Delta,n} e^{i \theta \int d^4 x \frac{1}{32 \pi^2} G \tilde G} \langle  n+\Delta \mid \OO  \mid n \rangle \\
&=& \sum_\Delta \int dA e^{i \int d^4 x \LL + \frac{\theta}{32 \pi^2} G \tilde G} \delta ( \Delta - \int d^4 x \frac{1}{32 \pi^2} G \tilde G) = \int dA e^{i \int d^4 x \LL + \frac{\theta}{32 \pi^2} G \tilde G} .
\eea
Hence choosing the $\theta$ vacua in the Hamiltonian is completely equivalent to having the $\theta$ term in the Lagrangian.

Thus we have finally resolved the two puzzles presented in the introduction.  The first question was why did $\theta$ have an effect when it does not appear in the Hamiltonian.  The answer is that it appears instead as a super-selection rule upon the states we are considering.  The second puzzle was the naive estimate that $\int d^4x G \tilde G/32 \pi^2$ should be zero making the action independent of $\theta$.  We showed that with an instanton background, it does not vanish and is instead an integer.  Thus $\theta$ does appear in the action and does have an effect.

\section{Non-axion solutions to the Strong CP problem} \label{Sec: not axion}

In this section, we briefly discuss non-axion solutions to the Strong CP problem.  Axion-type solutions will be explored in their own section.  While presenting these solutions, I will make a few sociological statements based on my own biases.  I strongly encourage students to make their own rankings independent of what other people consider interesting, since nature doesn't care what we think so
we should explore all options.

\subsection{The massless up quark}

The massless up quark is the simplest solution to the Strong CP problem~\cite{'tHooft:1976up} but has been experimentally ruled out~\cite{Beringer:1900zz,Dine:2014dga,Aoki:2016frl}.
The neutron dipole moment is a CP-odd quantity.  Taking the 2-flavor QCD Lagrangian, the only CP-odd constants
are $\theta$, $\theta_u$ and $\theta_d$.  As mentioned before, various anomalous symmetries can be used
to shift these constants around :
\bea
u \rightarrow e^{i \alpha} u , \qquad \theta_u \rightarrow \theta_u - \alpha , \qquad \theta \rightarrow \theta + \alpha
\eea
and
\bea
d \rightarrow e^{i \alpha} d , \qquad \theta_d \rightarrow \theta_d - \alpha , \qquad \theta \rightarrow \theta + \alpha.
\eea
Physical quantities such as the neutron eDM cannot depend on field redefinitions and must be invariant under these anomalous symmetries.  The only physical quantity is thus
\bea
\overline \theta = \theta + \theta_u + \theta_d
\eea

However, if the up quark is massless, there are only two CP-violating terms in the Lagrangian, $\theta_d$ and $\theta$, and the new 
anomalous symmetries are instead
\bea
u \rightarrow e^{i \alpha} u , \qquad \theta \rightarrow \theta + \alpha
\eea
and
\bea
d \rightarrow e^{i \alpha} d , \qquad \theta_d \rightarrow \theta_d - \alpha , \qquad \theta \rightarrow \theta + \alpha .
\eea
With these anomalous symmetries, it is impossible to construct a quantity that does not transform under the anomalous symmetries,
and thus it is impossible to write down a neutron eDM.  Put another way, if the up quark is massless, it is possible using field redefinitions to set every CP violating parameter to zero and the theory has a CP symmetry.

An alternate way of understanding the massless up quark solution is to calculate the neutron eDM explicitly using Eq.~\ref{Eq: neutronedm}, and to note that it is proportional to $\mu = m_u m_d/(m_u + m_d)$.  If
\bea
m_u < 10^{-10} m_d
\eea
then the neutron eDM will satisfy the current bounds.  
It is amusing to note that the massless up quark is an axion solution to the Strong CP problem where the already discovered $\eta'$ particle is the axion.
However, current lattice results show that the mass of the up quark is non-zero, ruling out this solution.  There have been a few attempts to build models that are similar in spirit to the massless up quark~\cite{Hook:2014cda,Agrawal:2017evu}, but other than that, the massless up quark solution is resting peacefully in its grave.

\subsection{RG running of $\overline \theta$}

In the following two solutions, RG running of $\overline \theta$ will be important.  In particular, these solutions will attempt to set $\overline \theta = 0$ at some RG scale and utilize the fact that in the SM, RG running of $\overline \theta$ occurs at 7-loops~\cite{Ellis:1978hq}.  If $\overline \theta$ is set to 0 at some high scale and the EFT to low energies is just the SM, then $\overline \theta$ will still be very small at low energies, thus it can be effectively ignored.

As no one in his or her right mind would try to perform a 7-loop computation, we will instead use symmetries to show that all 6-loop diagrams and below cannot generate any RG running.  The standard lore is that anything not forbidden by symmetries will be non-zero, so we say that at 7-loops there will probably be a non-vanishing correction.
The RG running of $\overline \theta$ comes from the only other CP-violating phase in the SM, $\theta_{CKM}$.  This can be seen by treating $\overline \theta$ and $\theta_{CKM}$ as spurions of the CP symmetry.  It must also respect the $SU(3)_Q \times SU(3)_{u^c} \times SU(3)_{d^c}$ flavor symmetries of the SM.  Thus,
\bea
\beta_{\overline \theta} = \text{arg  Tr} \prod Y_u Y_d Y_u^\dagger Y_d^\dagger .
\eea
A fun exercise is to write down a product of Yukawa matrices that respects the symmetries, but does not have a vanishing phase.  For example, the simplest product Tr $Y_u Y_u^\dagger$ has no phase.  It turns out that a non-vanishing phase requires
\bea \label{eq: index}
\beta_{\overline \theta} = g^2 \text{arg  Tr} Y_u^4 Y_d^4 Y_u^2 Y_d^2 ,
\eea
where we have included an extra factor of $g^2$ because $\overline \theta$ needs to see both CP and P violation.  It is left as an exercise for the reader to specify the non-trivial contractions of the indices in Eq.~\ref{eq: index}.  Twelve Yukawas and two gauge couplings mean that we need to write down a 7-loop contribution to see any RG running.

It is important to note that $\theta$ is a topological parameter that does not appear in perturbation theory.  Instead, it the phase of the quark masses that evolves with RG.

\subsection{Parity}

The first ``UV" solution to the Strong CP problem that we discuss is parity~\cite{Babu:1989rb,Barr:1991qx}\footnote{By UV, what we mean is that it sets $\overline \theta = 0$ at a scale larger than the QCD scale and then utilizes the small RG running to result in a small neutron eDM.}.  As mentioned above, if parity were a good symmetry of nature, then the
neutron eDM would be zero.  To see this in field theory language, we note that under parity
\bea
P : SU(2)_L \leftrightarrow SU(2)_R , \qquad Q_L \leftrightarrow Q_R^\dagger , \qquad H_L \leftrightarrow H_R^\dagger , \qquad L_L \leftrightarrow L_R^\dagger ,
\eea
where we have collected the right-handed quarks into $Q_R$, the right-handed leptons/neutrinos into $L_R$ and introduced a gauged $SU(2)_R$ and $H_R$ to make the electroweak sector parity invariant.
The $\theta$ term is P and CP odd and is forbidden by parity, while the Yukawas are of the form
\bea
\LL \supset \frac{y_u Q_L H_L Q_R H_R}{\Lambda_u} + \frac{y_d Q_L H_L^\dagger Q_R H_R^\dagger}{\Lambda_d} + h.c. .
\eea
The standard Yukawa matrices ($Y$) appear after the right-handed Higgs obtains its vev.  Under parity, the Yukawa matrices obey
\bea
Y_u = \frac{y_u v_R}{\Lambda_u} = Y_u^\dagger , \qquad Y_d = \frac{y_d v_R}{\Lambda_d} = Y_d^\dagger .
\eea
Hermitian matrices obey arg det $Y = 0$, but their individual elements can be complex.  Thus we see that the CKM matrix can have a non-trivial phase, but the neutron eDM, which is proportional to
\bea
\overline \theta = \theta + \text{arg det} Y_u + \text{arg det} Y_d ,
\eea
vanishes.

Parity can be broken softly by giving the right-handed Higgs a larger bare mass than the left-handed Higgs.  All of the 
nice properties of the parity solution are preserved with this type of breaking and the Strong CP problem is solved.  
Thus the simplest parity solution to the Strong CP problem is on par with the axion in its simplicity.

There are many models that move beyond the minimal parity solution~\cite{Mohapatra:1995xd,Kuchimanchi:1995rp,Mohapatra:1996vg,DAgnolo:2015uqq}, although they quickly run into new issues that need to be addressed.  The beauty and annoyance of parity-based solutions is that tree-level phases, such as the CKM phase, are allowed at tree level.  While the CKM phase doesn't effect the $\theta$ angle much due to the flavor structure of the SM, other tree-level phases can cause 1-loop effects on $\theta$.  Many models that UV complete the Yukawa couplings involve a bi-fundamental Higgs field.  Once this field is added, new CP-violating couplings are allowed, and there are again 1-loop contributions to $\overline \theta$ that need to be addressed.  Additionally, obtaining a large top Yukawa requires that $\Lambda_u \sim v_R$ so that the UV completion is not far away from $v_R$.
Due to these issues and other sociological factors, parity solutions are not as popular as the axion solution.

\subsection{CP}

Solutions to the Strong CP problem that utilize CP symmetry are typically called Nelson-Barr models~\cite{Nelson:1983zb,Barr:1984qx}.  They posit that CP is a good symmetry in the UV, so that both the CKM and $\theta$ angles vanish.  However, because the CKM is observed to be large, while $\overline \theta$ is observed to be small, these models are carefully constructed so that CP breaking gives a large CKM angle while corrections to $\overline \theta$ are small.

The simplest example uses new vector-like quarks, $q$ and $q^c$, with hyper charge $\pm 1/3$~\cite{Bento:1991ez,Dine:2015jga}.  Additionally, there are more than one complex scalars $\eta^a$ which obtain complex expectation values, breaking CP.  The tree-level Lagrangian under consideration is
\bea
\LL = \mu q q^c + Y^{ij} H Q_i d_j^c + A^{i a} \eta^a d_i^c q .
\eea
The $4\times4$ mass matrix for the quarks is then
\bea
M = \begin{pmatrix}
  \mu & A \eta  \\
  0  & m_d 
 \end{pmatrix} ,
\eea
where $m_d = Y v$ is the $3\times3$ down quark mass matrix.  It is simple to check that at tree level arg det $M = 0$,
while the CKM phase is non-zero and large if $\mu \lesssim A \eta$.  By fiddling with the size of various Yukawa couplings, loop-level corrections to $\overline \theta$ can be made small.

Aside from this simplest of CP-based solutions, there are also a plethora of fun things you can do when building models with CP symmetry, e.g. Refs~\cite{Dine:1993qm,Barr:1996wx,Hiller:2001qg,Vecchi:2014hpa} just to name a few.
Even more so than parity-based solutions, CP-based solutions are very fragile, as many coincidences of scales are needed for the CKM angle to be large.  Additionally, couplings such as $\eta q q^c$ and $H Q q^c$ must be forbidden or $\overline \theta$ appears at tree level.  For these reasons, CP-based solutions have fallen out of fashion.

\section{Axions} \label{Sec: axion}

Axions and their variants are by far the most popular solution to the Strong CP problem.  As such, I'll dedicate a whole section to describing axions and variations on the axion theme.  The terminology surrounding axions can be slightly annoying and confusing :
\begin{itemize}
  \item QCD Axion : Solves the Strong CP problem
  \item ALP : Does not solve the Strong CP problem
  \item Axion : Figure it out yourself
\end{itemize}
If the reader encounters the word ``axion", he/she will have to determine from the context whether it solves the Strong CP problem.

\subsection{The QCD axion}

After the massless up quark, the axion~\cite{Peccei:1977hh,Peccei:1977ur,Weinberg:1977ma,Wilczek:1977pj} is typically considered to be the simplest solution to the Strong CP problem,  though the minimal parity-based solution gives the axion EFT a run for its money.
The EFT of the axion is extremely simple and is the main reason for its popularity.  The EFT consists of a single new particle, the axion ($a$), and a single new coupling ($f_a$) :
\bea
\label{Eq: EFT}
\LL \supset \left (\frac{a}{f_a} + \theta \right ) \frac{1}{32 \pi^2} G \tilde G .
\eea
We have written both the $\theta$ term and the axion coupling to demonstrate a simple trick that shows how the axion couples.
As is apparent from this interaction, the axion obeys an anomalous symmetry
\bea
a \rightarrow a + \alpha f_a , \qquad \theta \rightarrow \theta - \alpha ,
\eea
which dictates how the axion can couple to particles.  For example, every non-derivative interaction of the axion can be obtained by observing that wherever we have a coupling $\theta$, we can replace it with $\theta + a/f_a$.  Derivative couplings are more complicated because $\partial \theta = 0$, so they are not accompanied by a corresponding $\theta$ coupling.

UV completions of the QCD axion will occasionally generate other couplings, such as
\bea
\LL \supset \frac{a}{f_B} \frac{1}{32 \pi^2} B \tilde B + \frac{a}{f_W} \frac{1}{32 \pi^2} W \tilde W .
\eea
Axions with these additional couplings are still called the QCD axion as long as the axion still has the coupling shown in Eq.~\ref{Eq: EFT}.  Due to the anomalous symmetry structure of the axion and the topological nature of the spurion $\theta$, these couplings must be there initially, or they are not generated by RG evolution~\footnote{See Ref.~\cite{Craig:2018kne} for a more detailed symmetry based analysis of the RG running of axion couplings.}.  The other couplings generated by RG evolution are derivative interactions with quarks :
\bea
\frac{\partial_\mu a}{f_Q} Q^\dagger \sigma^\mu Q .
\eea
Even if these couplings are zero at tree level, they are still generated by RG evolution~\cite{Bauer:2017ris}.

The whole point of introducing the axion was to hopefully solve the Strong CP problem, so let's show that the axion sets the neutron eDM to zero.  First we calculate the axion mass and expectation value.  We already calculated how the vacuum energy of QCD depended on $\theta$ in Eq.~\ref{Eq: thetapot}.  Using our trick from before, we find that the axion potential is
\bea
V = - m_\pi^2 f_\pi^2 \sqrt{1 - \frac{4 m_u m_d}{(m_u + m_d)^2} \sin^2 \left (\frac{a}{2 f_a} + \frac{\overline \theta}{2}  \right) } .
\eea
Thus, the axion vev is $\langle a \rangle = -\overline \theta f_a$.
We can calculate the neutron eDM using the same trick to find that 
\bea
d_n \propto \frac{a}{f_a} + \overline \theta = 0 .
\eea
Thus, once the axion relaxes to the minimum of its potential, it dynamically sets the neutron eDM to zero.
As claimed, the QCD axion solves the Strong CP problem.

Because all of the CP breaking in QCD is dictated by the spurion $\overline \theta$, you can quickly convince yourself that all higher-order corrections to the axion potential coming from QCD do not shift the axion vev.  However, once you include all three generations of quarks, the CKM matrix has a CP-violating phase that can break CP and move the axion potential away from having the neutron eDM $= 0$.  This effect has been estimated
to be of the size $\langle a/f_a \rangle - \overline \theta \sim 10^{-18} - 10^{-20}$~\cite{Georgi:1986kr}.  It is of some comfort that in the asymptotic future, physicists will be able to test the axion solution to the Strong CP problem regardless of whether they ever find the axion particle itself.

\subsection{The axion quality problem}

The axion quality problem relates to the fact that while there is a very well-defined anomalous symmetry associated with the axion coupling shown in Eq.~\ref{Eq: EFT}, there is no symmetry associated with it.  
This lack of proper symmetry properties generates two issues that together are called the axion quality problem~\cite{Kamionkowski:1992mf,Barr:1992qq,Ghigna:1992iv} :

\begin{enumerate}
  \item EFTs are built by specifing the particle
content and symmetries of the problem, and then writing down every coupling allowed by symmetry.  Because the axion has no symmetry properties, there is no way to form the axion coupling in Eq.~\ref{Eq: EFT} without also including a host of other couplings.
  \item Quantum gravity is believed to break all symmetries that aren't gauged.  Thus, even if one imposes the anomalous symmetry, gravitational effects will break it and the axion will obtain a separate mass term that is not centered around a zero neutron eDM, which reintroduces the problem.
\end{enumerate}

To see the axion quality problem in action, we consider the simplest UV completion of the axion\footnote{The two most well-known UV completions of the axion are called KSVZ~\cite{Kim:1979if,Shifman:1979if} and DFSZ~\cite{Dine:1981rt,Zhitnitsky:1980tq}.}.  We have a complex scalar $\Phi$ with an approximate $U(1)$ symmetry, of which the axion is the pseudo-Goldstone boson.  This $U(1)$ symmetry is traditionally called the $U(1)_{PQ}$ symmetry.  In addition to $\Phi$, we have an additional pair of vector-like quarks $q$ and $q^c$ that are only charged under QCD.
The Lagrangian is
\bea
V = - m^2 \Phi \Phi^\dagger + \lambda (\Phi \Phi^\dagger )^2 + y \Phi q q^c + h.c.
\eea
$\Phi$ obtains an expectation value
\bea
\Phi = (f_a+r ) e^{i a/f_a} .
\eea
We can integrate out the radial mode $r$ and the now heavy quarks $q$ and $q^c$.  The resulting IR theory is that
of the axion.  The coupling of the axion in Eq.~\ref{Eq: EFT} can be seen to arise via the anomalous field redefinition
\bea
q' = q e^{i a/f_a} ,
\eea
which makes the axion appear in the $G \tilde G$ coupling due to the anomaly.

We can now see the axion quality problem in this context.  The first issue is that the $U(1)$ symmetry imposed on $\Phi$ and $q$ is an anomalous symmetry and thus by the standard rules of an EFT, we cannot forbid the couplings $\epsilon q q^c$ and $\epsilon^2 \Phi^2$, which break the anomalous $U(1)_{PQ}$ symmetry.  The second issue is that since gravity breaks this anomalous symmetry, we can expect higher dimensional operators of the form
\bea
V \sim \frac{\Phi^n}{M_p^{n-4}} .
\eea
As we will describe in the following section, the experimental constraints on the axion require that $f_a \gtrsim 10^8$ GeV, and there is a preferred region around $f_a \sim 10^{12}$ GeV for dark matter reasons~\cite{Abbott:1982af,Dine:1982ah,Preskill:1982cy}.  As such, we will set $f_a = 10^{12}$ GeV and see how severe the axion quality problem is.

The first issue is how small the coupling $m^2 \Phi^2$ needs to be.  This term generates a potential for the axion that is 
\bea
V \sim \epsilon^2 f_a^2 \cos \left ( \frac{a}{f_a} + \phi \right ) .
\eea
Requiring this potential to be small enough that the axion still solves the Strong CP problem results in $\epsilon \lesssim 10^{-19} GeV$.  Thus the anomalous symmetry must be extremely good.  The Planck-suppressed operators give a potential of the form
\bea
V \sim \frac{f_a^n}{M_p^{n-4}} \cos \left ( \frac{a}{f_a} + \phi_n \right ) .
\eea
Again, requiring that this potential does not shift the minimum away from $\overline \theta$ by more than $10^{-10}$ gives the constraint that $n \gtrsim 14$.  We would need to prevent Planck-suppressed operators to a ridiculously high order to solve the Strong CP problem!  
This toy UV completion highlights how serious this axion quality problem is and why significant effort has been devoted to it over the years.

\subsection{Solving the axion quality problem}

People have developed many ways to solve the axion quality problem.  These broadly fall into the categories of 
theories where the $U(1)_{PQ}$ is an accidental symmetry, theories where the $U(1)_{PQ}$ comes from 5D gauge symmetries, and examples from string theory~\cite{Witten:1984dg}.  I will discuss the first two approaches in this section and skip the string theory examples as they require a lot more background that is beyond the scope of these lectures.  None of these approaches comes close to the elegance of the axion EFT. 
The fact that such complicated models are needed to justify an elegant EFT is the dirt that typically gets swept under the rug in axion discussions.

Accidental $U(1)_{PQ}$ symmetries can come from discrete symmetries.  For example, as shown in the previous subsection, a $\mathbb{Z}_{14}$ discrete symmetry acting on $\Phi$ could prevent all of the dangerous higher-dimensional operators.  More plausibly, the accidental $U(1)_{PQ}$ symmetry could result from chiral gauge theories in much the same way that $U(1)_{B-L}$ is an accidental symmetry of the renormalizable SM. 
A model that realizes this has four sets of quarks, $Q_{1,2,3,4}$, as well as two new confining gauge groups, $SU(N)$ and $SU(M)$~\cite{Randall:1992ut}.  The pseudo Goldstone boson whose breaking gives the axion is the $U(1)_{PQ}$ shown below :
\begin{center}
\bea
\begin{tabular}{c|ccc|c}
&$SU(3)_c$&$SU(N)$&$SU(M)$&$U(1)_{PQ}$\\
\hline
&&\\[-8pt]
$Q_1$ & $\fund$ & $\fund$ & $\fund$  &  1 \\
$3 \times Q_2$ &  & $\fund$ & $\fund$  & -1 \\
$M \times Q_3$ & $\overline \fund$ & $\overline \fund$ &   & 1  \\
$3M \times Q_4$ & & $\overline \fund$ &   & -1 \\
\end{tabular}
\eea
\end{center}
where we have indicated that there are three copies of the quark $Q_2$, M copies of $Q_3$ and 3M copies of $Q_4$.
After some amount of work, it can be shown that after $SU(N) \times SU(M)$ confines there is a light pseudo Goldstone boson that is the axion.  The lowest-dimensional operator we can write down that violates this accidental $U(1)_{PQ}$ symmetry is
\bea
\LL \supset \frac{Q_2^M Q_4^M}{M_p^{3M - 4}} \sim \frac{f_a^{3M}}{M_p^{3M - 4}} \cos \left ( \frac{a}{f_a} + \phi \right ) .
\eea
By taking the gauge group $M \geq 5$, we can suppress higher-dimensional operators enough that the axion still solves the Strong CP problem.  Of course this is not the only model that does solves the axion quality factor in this manner, see e.g. Ref.~\cite{Lillard:2018fdt}.

On the other hand, the extra-dimensional approach solves the problem with Planck-scale physics by promoting the axion shift symmetry to a gauge symmetry in the UV.  This can be done in several ways.  One way is to make the axion the $A_5$ of a 5D gauge field~\cite{ArkaniHamed:2003wu} such that
\bea
a/f_a = \int dy A_5 .
\eea
Thus at high energies, the axion is protected from corrections due to 5D gauge invariance.  The coupling to gluons can result from a 5D Chern-Simons interaction~\cite{Choi:2003wr}.  Another 5D approach is to use a gauged $U(1)_{PQ}$ in the bulk.  Accidental anomalous $U(1)_{PQ}$ symmetries can result if the fermions that canel anomalies are localized on different branes.  If there are no $U(1)_{PQ}$ charged particles in the bulk, then the accidental PQ symmetries can be of very high quality~\cite{Cheng:2001ys,Izawa:2002qk}.  All of these 5D solutions can be dimensionally deconstructed into 4D solutions, but the 5D versions motivate the particular structure of the theories.

\subsection{Variations of the QCD axion}

The EFT of the QCD axion is very elegant and simple.  Its coupling to the gluons and its mass are intrinsically tied together, while the couplings to the fermions and photons are model dependent.  As with any simple and elegant theory, people have started to push various boundaries.  
Two basic variations of the QCD axion have been explored.  The first results in larger-than-expected couplings to fermions and photons.  The second breaks the mass to neutron coupling relationship.

\subsubsection{Large fermion and photon couplings}

The coupling of the QCD axion to fermions, $f_Q$, and photons, $g_{a \gamma \gamma}$, can be much larger or smaller than $f_a$,
\bea
\frac{g_{a \gamma \gamma} a}{4} F \tilde F \qquad \frac{\partial_\mu a}{f_Q} Q^\dagger \sigma^\mu Q .
\eea
It turns out to be very difficult to make $f_Q$ much smaller than $f_a$, as RG evolution tends to bring $f_Q$ to within a loop factor of $f_a$.  Similarly, it is very difficult to make $g_{a \gamma \gamma}$ much smaller than $1/f_a$ because mixing with the pions couples the axion to photons~\cite{Srednicki:1985xd,Georgi:1986df}.  
\bea
g_{a \gamma \gamma} = \frac{\alpha}{2 \pi} \left [ \frac{1}{f_\text{UV}} - \frac{1.92(4)}{f_a} \right ] ,
\eea
where $f_\text{UV}$ is the model-dependent coupling and the second term comes from QCD.
Photon couplings parametrically smaller than $f_a$ are only achievable if the two terms cancel each other.  This cancellation accidentally happens for some GUTs where $f_\text{UV} = f_a/2$~\cite{Giudice:2012zp}, and can happen if one chooses Casmirs of the UV quarks to give $f_\text{UV} \sim f_a/1.92$~\cite{Kaplan:1985dv,DiLuzio:2016sbl,DiLuzio:2017pfr}.  A way of obtaining parametrically small $g_{a \gamma \gamma}$ without fine tuning is not known.

There has been a recent jump in popularity of trying to make the axion coupling to fermions or photons larger than $1/f_a$.  For example, the simplest way to increase the axion-photon coupling is to dial up the electric charge of the quarks ($q,q^c$) that we integrate out to get the axion-gluon coupling.  This approach can boost the coupling to photons until $g_{a \gamma \gamma} \sim 1/f_a$ before the Landau pole of $U(1)_Y$ is too low for the EFT to make any sense.

More recent attempts have all centered around using multiple axions and mixing them in such a way that the coupling to gluons is small while the coupling to fermions and photons is large~\cite{Farina:2016tgd,Agrawal:2017cmd}.  Although these approaches have been around for a long time~\cite{Dvali:2007hz,Choi:2014rja}, a move of marketing genius has recently renamed them as clockwork models~\cite{Kaplan:2015fuy,Choi:2015fiu}.  The simple version is
\bea
\frac{b}{f_b} \frac{\theta g_s^2}{32 \pi^2} G \tilde G + \left( \frac{N b}{f_b} + \frac{a}{f_a} \right)  \frac{\theta g_s'^2}{32 \pi^2} G' \tilde G' + \frac{a}{f_a} \frac{e^2}{32 \pi^2} F \tilde F ,
\eea
where $G'$ is a new confining gauge group, $N$ is an integer coming from the group theory of the UV completion, and $a$ and $b$ are pseudo-scalars.  Assuming that $G'$ confines at a high scale, integrating out the linear combination $N a/f_a + b/f_b$ and defining $f = N f_a$, we get the IR Lagrangian
\bea
\frac{a}{f} \frac{\theta g_s^2}{32 \pi^2} G \tilde G + \frac{N a}{f} \frac{e^2}{32 \pi^2} F \tilde F .
\eea
$N$ can be taken to be large, giving us a parametrically enhanced coupling of the axion to the photons.  Alternatively, $N$ can be kept small while the number of particles ($a$, $b$, $c$, ...) is taken to be large.

\subsubsection{Changing the axion mass - neutron coupling relation}

The other way of playing with the axion is to change the relationship between the coupling of the axion to the neutron eDM and the mass of the axion.  Both of these terms come from the $G \tilde G$ coupling so that the relationship is pretty tight in the usual case.  To change it, another contribution to the axion mass must be present.  This extra contribution has to be centered around where the neutron eDM is zero, otherwise the axion fails to solve the Strong CP problem.

There are two general approaches to make the axion heavier than expected given its coupling to the neutron eDM.  The first is to introduce a new confining gauge group to which the axion couples~\cite{Rubakov:1997vp,Berezhiani:2000gh,Hook:2014cda,Fukuda:2015ana,Blinov:2016kte,Dimopoulos:2016lvn} :
\bea
\left( \frac{a}{f_a} + \theta \right)  \frac{g_s^2}{32 \pi^2} G \tilde G +  \left( \frac{a}{f_a} + \theta' \right)  \frac{g_s'^2}{32 \pi^2} G' \tilde G' .
\eea
In order to still solve the Strong CP problem, we need $\overline \theta \approx \overline \theta'$ up to $10^{-10}$.  
A discrete symmetry, usually $\mathbb{Z}_2$, is used to make these angles identical.  If the $\mathbb{Z}_2$ symmetry is only very softly broken, e.g. the Higgs and mirror Higgs have different masses, then both $\overline \theta$ will remain equal due to the small RG running of $\overline \theta$ in the SM.  The axion will still solve the Strong CP problem while the mass will be larger than expected from the neutron eDM coupling.

The second way to increase the axion mass is to use QCD itself to give the axion a second contribution to its mass~\cite{Holdom:1982ex,Holdom:1985vx,Dine:1986bg,Flynn:1987rs,Choi:1988sy,Choi:1998ep,Agrawal:2017ksf} using UV instantons.  As argued in the previous section, there exist instanton configurations that are sensitive to $\theta$ and thus contribute to the axion potential.  Evaluating $\int d^4 x G^2$ in the background of the instanton shows that the instanton potential scales in the UV as
\bea
V(a) \sim e^{- \frac{8 \pi^2}{g^2}} \prod_f y_f \cos \left ( \frac{a}{f_a} - \overline \theta \right ) .
\eea
Since QCD is perturbative in the UV, these extra contributions are usually negligible.  However, if QCD becomes strong in the UV, then they can give the axion a mass that is larger than otherwise expected.  We can accomplish this by adding additional matter so QCD is not asymptotically free, or by utilizing Higgsing.

Finally, there is only a single model that can make the axion lighter than expected~\cite{Hook:2018jle}.  Because the neutron eDM coupling to the axion is a non-derivative interaction, the axion has a hierarchy problem that is resolved by QCD dynamics.  If the axion is to be lighter than expected, one not only has to solve the hierarchy problem, but also ensure that $\overline \theta = 0$ is still the minimum.  The model which does this utilizes a discrete $\mathbb{Z}_N$ symmetry.  The axion non-linearly realizes the $\mathbb{Z}_N$ symmetry and interacts with N copies of QCD given by
\bea
\LL = \sum_k \left ( \frac{a}{f} + \frac{2 \pi k}{N} + \theta \right ) G_k \tilde G_k .
\eea
Surprisingly, a numerical check shows that adding all these different contributions to the axion potential not only retains a minimum around $\overline \theta=0$ if $N$ is odd, but also  makes the axion mass exponentially lighter :
\bea
\frac{m_a(N)}{m_a(N=1)} \sim \frac{4}{2^{N/2}} .
\eea
This is because the axion potential in this case can be related to a Riemann sum that is known to converge exponentially fast for periodic analytic potentials.

\subsection{ALPs}

Unless one accepts
complex models, $g_{a \gamma \gamma} \sim 1/f_a$ for the QCD axion.  Thus the mass and the coupling to the photon, the easiest of the axion couplings to probe, are related, though only at the tilde level.  Many experiments have been devoted to search for the QCD axion through this coupling, even though $g_{a \gamma \gamma}$ is completely unrelated to the Strong CP problem.
Axion-like particles (ALPs) address the question : If we're looking for a particle whose couplings have nothing to do with the Strong CP problem, then why does the particle have to solve the Strong CP problem?  ALPs are particles which have the Lagrangian
\bea
\LL = \frac{1}{2} m_a^2 a^2 + \frac{g_{a \gamma \gamma} a}{4 f} F \tilde F .
\eea
This mass may or may not come from the confinement of another gauge group.
Because there is no coupling to gluons, the mass and the coupling to photons are completely independent from each other.
There are string theory motivations for why these new particles may exist~\cite{Arvanitaki:2009fg}.  However, having seen so many pretty
models crash and burn, perhaps the strongest motivation for looking for ALPs is that due to recent technological 
advances, we can now cover many orders of magnitudes of new parameter space, so we should look and see what 
we find.

\section{Axion/ALP dark matter} \label{Sec: dark matter}

One of the appealing qualities of the axion is that it can also be dark matter, solving two problems at the same time~\cite{Abbott:1982af,Dine:1982ah,Preskill:1982cy}.
The axion can obtain its DM abundance from either the misalignment mechanism or topological defects.  In this 
section, we review how to estimate the number abundance of axions from these various processes.

The misalignment mechanism operates when PQ symmetry is broken during inflation and is never restored after 
inflation\footnote{This is a slight lie as misalignment is also present when PQ symmetry is restored.  In this case, the initial angle is random and can be averaged over all Hubble patches.  The inhomogeneity of the initial axion angle leads to different phenomenology.}.  Inflation results in the same initial conditions being seen everywhere.  The topological mechanism 
operates when PQ symmetry is restored either during inflation or after inflation.

\subsection{Misalignment : ALP Dark Matter}

We first study the case of ALP dark matter.  In a Friedmann-Robertson-Walker (FRW) universe, the equation of motion for 
ALP dark matter is
\bea
\label{Eq: eom-ALP}
\ddot a + 3 H \dot a + m_a^2 a = 0 .
\eea
People typically treat a radiation-dominated universe where $H = 1/2t$.  We assume that the axion has an
initial field value $a = a_0$.  In the early universe, Eq.~\ref{Eq: eom-ALP} describes an over-damped 
harmonic oscillator.  Thus we approximate 
\bea
\label{Eq: early ALP}
a = a_0 , \qquad H \gg m_a .
\eea  
In the late-time universe,
Eq.~\ref{Eq: eom-ALP} describes an under-damped harmonic oscillator.  Using the WKB approximation, we find in the far future that
\bea
\label{Eq: late ALP}
a =  (\frac{R(H = m_a)}{R(t)})^{3/2} A_0 \cos m_a t , \qquad H \ll m_a ,
\eea
where $R$ is the scale factor.  When estimating the ALP number abundance, we take Eq.~\ref{Eq: early ALP} and Eq.~\ref{Eq: late ALP} to apply when $3 H > m_a$ and $3 H < m_a$ respectively, ignoring the crossover regime.
Using these approximations, we have $A_0 \approx a_0$.

We note that at late time, the axion behaves like cold non-relativistic matter so that its energy density red-shifts as
\bea
\rho(t) = \rho(H= m_a) (\frac{R(H = m_a)}{R(t)})^{3} .
\eea
Additionally, its energy is found via Fourier transformation to be equal to its mass.  Thus, we have a production mechanism for cold dark matter (CDM) that works even when the particle is much much lighter than the keV scale.  This is an impressive feat, as usually dark matter with a mass below a keV behaves like hot dark matter.

The energy density of the SM and ALP when the ALP starts acting like CDM is
\bea
\rho_{SM} \sim H^2 M_p^2 \sim m_a^2 M_p^2 , \qquad \rho_{a} \sim m_a^2 a_0^2.
\eea
Requiring that the ALP accounts for all of DM ($\rho_{DM} \sim  T^3$ eV) gives
\bea
a_0^2 \sim \text{eV} \, \frac{M_p^{3/2}}{m_a^{1/2}} .
\eea
As long as the initial value of the ALP is this value, then it will make up all of DM.

\subsection{Misalignment : Axion DM}

The estimate for axion DM is very similar to that of ALP DM, but with a critical difference.  The
axion mass changes with time.  As a result, one needs to be careful with the lack of energy conservation.
 Let us first estimate the temperature dependence of the axion mass.  

The axion mass comes from
thermal instantons~\cite{Gross:1980br} and we will use dimensional analysis to estimate the result.  It turns out that we will be interested
in the axion mass around temperatures between 100 MeV and 1 GeV.  In this range, QCD acts like  three-flavor QCD.  As with the massless up quark solution to the Strong CP problem, if any quark mass goes to zero then $\overline \theta$ becomes unphysical.  Thus the axion potential must be proportional to the product of the masses.  The remaining dimensions can be made up using the temperature T.
The next factor is the exponential suppression of the potential $e^{- 8 \pi^2/g_3^2(T)}$ present for all instanton calculations.  The QCD coupling $g_3$ can be evaluated using the 1-loop expression for the RG running.  The end result for the potential is
\bea
V \sim m_u m_d m_s T e^{- 8 \pi^2/g_3^2(T)} \cos \left ( \frac{a}{f_a} + \overline \theta \right ) \sim m_u m_d m_s \frac{\Lambda^9}{T^8} \cos \left ( \frac{a}{f_a} + \overline \theta \right ) .
\eea
As a result, the temperature dependence of the axion is
\bea
m_a(T)^2 \sim \frac{m_u m_d m_s}{f_a^2} \frac{\Lambda^9}{T^8} .
\eea

Finally, before we start estimating the axion DM abundance, we roughly specify the initial conditions for the axion.  We take $a_0 = \theta_0 f_a$.
Because the axion is a periodic field, we work with the misalignment angle $\theta_0$.  The generic assumption people make is that $\theta_0 \sim \OO(1)$, simply because of our affinity for $\OO(1)$ numbers.
In some cases, the initial angle can be estimated in an inflationary context.
Inflation will kick the axion expectation value around by an amount $\sim H$ every Hubble time in the form of a random walk.  This leads to inflation populating every value of $\theta_0$.  Depending on the choice of
how to deal with the measure problem, this can even lead to predictions for the value of $\theta_0$.

Another effect of inflation kicking around the axion expectation value is inhomogeneities between various Hubble patches.  As a result, different Hubble patches will have different dark matter densities, leading to well-known isocurvature constraints if $H/f_a$ isn't small enough~\cite{Axenides:1983hj,Linde:1985yf,Seckel:1985tj,Lyth:1989pb,Turner:1990uz}.

We are now in a position to estimate the axion DM abundance.  The equations of motion for the axion are
\bea
\label{Eq: eom-axion}
\ddot a + 3 H \dot a + m_a^2(T) a = 0 .
\eea
In the early universe, we do the same thing as before and estimate
\bea
\label{Eq: early}
a = \theta_0 f_a , \qquad H > m_a(T).
\eea  
Again we will completely ignore transition regions.  At a critical temperature, $T_c$,
\bea
H(T_c) = m_a(T_c) .
\eea
At this point, the axion starts to oscillate.  After the axion starts to oscillate, we again want to apply the WKB
approximation, which can be used as long as
\bea
H \ll m_a , \qquad \dot m_a \ll m_a^2 .
\eea
Both of these are saturated around $T_c$ and as before, we trade the $\ll$ for a $<$.
With a changing mass, the WKB approximation tells us that number density not energy density is conserved and the expansion of the universe gives the standard volume dilution.  Thus we find that
\bea
n_a(T) \sim m_a(T) a(T)^2 \approx n_a(T_c) (\frac{R(T_c)}{R(T)})^{3} , \qquad T < T_c .
\eea
Once the axion mass stops changing around 100 MeV, energy also dilutes away like CDM.

Putting everything together, we have 
\bea
a = \theta_0 f_a , \qquad T > T_c ,
\eea  
and
\bea
a = \theta_0 f_a \sqrt{\frac{m_a(T_c)}{m_a(T)}} (\frac{R(T_c)}{R(T)})^{3/2}  \cos m_a t , \qquad T < T_c.
\eea

We can now compare the axion and dark matter energy densities at $T \sim \Lambda_{QCD} \sim 100$ MeV, where conservation
of energy in the axions becomes a good approximation :
\bea
\label{Eq: oofff}
\rho_a \sim \theta_0^2 \Lambda_{QCD}^4 \frac{m_a(T_c)}{m_a} \left ( \frac{\Lambda_{QCD}}{T_c} \right )^{3}
\sim \theta_0^2 \Lambda_{QCD}^4 \frac{f_a \Lambda_{QCD}}{T_c M_p} \sim \rho_{DM} \sim \text{eV} \, \Lambda_{QCD}^3 ,
\eea
where we have used that $m_a^2 f_a^2 \sim \Lambda_{QCD}^4$.  
Solving Eq.~\ref{Eq: oofff}, we find that $T_c \sim$ GeV and $f_a \sim 10^{11}$ GeV.  Using the instanton approximation with proper $\OO(1)$ numbers and solving the differential equations exactly gives $f_a \sim 2 \times 10^{11}$ GeV~\cite{diCortona:2015ldu}.  You'll sometimes hear people quote a number closer to $f_a \sim 10^{12}$ GeV because extrapolated lattice results~\cite{Buchoff:2013nra} suggest a slightly larger value of $f_a$ (lattice simulations have not been able to simulate physical values of quark masses yet).

\subsection{Topological production of axions}

After this relatively in-depth review of the misalignment mechanism, I will now give an unfairly brief review of the 
topological production of axions~\cite{Vilenkin:1982ks,Sikivie:1982qv,Vilenkin:1984ib,Davis:1985pt,Davis:1986xc}.  This is partly because of my own ignorance on the subject and partly because this field has not yet reached a clear consensus on the results~\cite{Kawasaki:2014sqa,Klaer:2017ond}.  It is prohibitively difficult to simulate realistic values of parameters, so all groups need to extrapolate their results.

PQ symmetry is a $U(1)$ symmetry that is only broken by QCD.  It turns out that it has strings~\cite{Kibble:1976sj,Kibble:1980mv,Vilenkin:1981kz} because $U(1)$ is a circle, and if we look asymptotically far away from a string in the transverse directions, space is also a circle.  As mentioned before, when mapping a circle to a circle, there is an integer winding number.  The vacuum has 0 winding number, whereas objects called strings have non-zero winding number.  

After PQ symmetry breaking, each Hubble patch will settle down to its own value of the axion.  As various patches come into contact, some might form a non-trivial winding number by chance.  Thus on average, one string per Hubble volume is created.  As the universe expands, more strings come into contact with each other and relax towards their ground state by closing loops and radiating axions.  The expectation is that approximately $\OO(1)$ number of strings is left in each Hubble volume at all times~\cite{Albrecht:1984xv,Bennett:1987vf,Allen:1990tv,Vincent:1996rb,Martins:2000cs,Gorghetto:2018myk,Kawasaki:2018bzv}.  The axions emitted as the strings merge and straighten out are mostly emitted with very low energy, $\OO(H)$, and have the potential to be dark matter~\cite{Vilenkin:1986ku,Davis:1989nj,Dabholkar:1989ju,Battye:1993jv,Battye:1994au,Yamaguchi:1998gx,Yamaguchi:1999yp,Hagmann:2000ja}.
Current estimates are that if $f_a \sim 10^{11}$ GeV, with about an order of magnitude uncertainty, then axions can be dark matter via topological production.

The ultimate fate of the string depends on whether there is a preserved subgroup of the PQ symmetry.  If the interactions with QCD break the $U(1)_{PQ}$ completely, then the strings decay.  If the interactions with QCD preserve a discrete subgroup of the $U(1)_{PQ}$, then some strings are stable and over-close the universe.

\subsection{Variations of dark matter axions}

The energy density of axions is roughly
\bea
\Omega_a h^2 = 0.01 \theta_0^2 \left ( \frac{f_a}{10^{11} \, \text{GeV} }\right)^{1.19}
\eea
when $f_a$ is sub-Planckian.  For $f_a \sim M_p/10$, the axion starts to oscillate after its mass losses its temperature dependence and the results scale differently.  Usually, people are interested in a single $f_a \sim $ few $\times 10^{11}$ GeV where $\theta_0 \sim 1$.  There are several directions for generalizing dark matter axions.

One variation is to arrange for axions to be dark matter even when $f_a \gtrsim 10^{11}$ GeV.  This can occur if $\theta_0 \lesssim 1$ because if the initial angle is smaller, then $f_a$ can be larger.  One of the major arguments for this is anthropics~\cite{Linde:1987bx,Tegmark:2005dy,Ubaldi:2008nf}.  Since inflation populates all values of $\theta_0$ and the existence of life as we know it might correlate with small $\theta_0$.   Axions can also be DM via an entropy dump approach~\cite{Dine:1982ah,Steinhardt:1983ia,Lazarides:1990xp,Kawasaki:1995vt}.  An entropy dump heats up the SM while leaving DM alone, so that the relative energy density in DM decreases.  Finally, the last approach I'll mention is to take energy out of the DM and dump it into another sector via particle production~\cite{Agrawal:2017eqm,Kitajima:2017peg}.

The next variation is to have axion dark matter even when $f_a \lesssim 10^{11}$ GeV.  In this case, there is not enough dark matter in axions and we need a way to generate even more cold axions.  The simplest way is to tune $\theta \sim \pi$~\cite{Turner:1985si,Lyth:1991ub,Visinelli:2009zm}.  Because the axion potential is a cosine, sitting near $\pi$ means that the axion takes a much longer time than $1/m_a$ to oscillate, so it starts to behave as cold dark matter much later.  One can also make lower-$f_a$ dark matter axions using topological defects and fine tuning~\cite{Kawasaki:2014sqa,Visinelli:2009kt,Hiramatsu:2010yn,Hiramatsu:2012sc}.  Another way to get small $f_a$ axion DM is parametric resonance~\cite{Co:2017mop}, which is essentially Bose-enhanced decays.  If another scalar has enough energy to be dark matter, its Bose-enhanced decays into axions could result in axions being dark matter.

The last variation on dark matter axions explores how they show up in the galaxy today.  In Sec.~\ref{Sec: DMaxion}, I will cover the typical assumptions for how axion DM behave locally.  Additionally, people sometimes discuss formation mechanisms and signatures of axion miniclusters~\cite{Hogan:1988mp,Kolb:1993zz,Kolb:1995bu}, which are 
typically AU in size and much lighter than the mass of a star.  Note that this is still 10 orders of magnitude more dense 
than dark matter in the galaxy.  
Another possibility people play with is Bose/axion stars~\cite{Schiappacasse:2017ham,Visinelli:2017ooc}.  These objects are even denser than miniclusters.

\section{Experimental probes of the Axion} \label{Sec: experiments}

Recently there has been a boom in the number of proposed experiments to look for the axion.
As there already exist reviews that go into excruciating detail~\cite{Graham:2015ouw,Irastorza:2018dyq},
I will give the overall idea of how each search strategy works and the region of parameter space that it probes.
The various searches will be categorized by if the axion is dark matter.

\subsection{DM independent searches}

\subsubsection{Rare meson decays}

If $f_a \lesssim 10^4$ GeV and $m_a \lesssim 100$ MeV, the axion/ALP can be probed by rare meson 
decays~\cite{Peccei:1988ci,Turner:1989vc,Alves:2017avw,PDG}.  Because the QCD axion mixes with some of the neutral pions, it can appear in some of the rare meson decays.  Even if it doesn't mix with the pions, \`a la ALPs, it can still appear in decays through its coupling to gauge bosons~\cite{Izaguirre:2016dfi}.

\subsubsection{Stellar cooling}

If $f_a \lesssim 10^{7-8}$ GeV and $m_a \lesssim 100$ keV, then the axion/ALP can be produced in stars~\cite{Raffelt:2006cw}.  We understand energy transport and cooling of stars well enough to impose constraints on whether the axions escape, or even just transport energy from the core to the outside of the star.  The upper bound on $m_a$ of 100 keV is due to the temperature in the center of stars.

Axions in stars are produced via a variety of mechanisms depending on the couplings.  One method is Bremsstrahlung $ e + N \rightarrow e + N + a$, in which the axion, $a$, is radiated from the electron and $N$ is a nucleus, usually ionized so that it has a large charge.  Another production mechanism is Primakoff radiation $\gamma + e \rightarrow e + a$, where the photon is converted into an axion via the axion-photon coupling.

\subsubsection{Supernova}

Supernova bounds apply when $10^6$ GeV $\lesssim f_a \lesssim 10^8$ GeV and $m_a \lesssim 100$ MeV~\cite{Raffelt:2006cw}.  As before, the mass cutoff is due to the temperature of the supernova being 10s of MeV.  Supernovae can be used to constrain axions by requiring that the axions produced do not carry away energy equal to the total energy in the neutrinos emitted by the supernova.  There is only a range of $f_a$ excluded by supernovae because eventually the axions become too strongly coupled to escape and no longer provide a good cooling mechanism.  Additional bounds come from axion production in supernovae and subsequent conversion into photons in the galactic magnetic fields~\cite{Payez:2014xsa}.

Due to the high near-nuclear densities in the center of supernovae, the dominant production of axions comes from nuclear Bremsstrahlung $n + n \rightarrow n + n + a$, where the axion is radiated by a neutron and the neutrons scatter via a pion exchange.  It is important to note that we do not quite understand supernovae and that the bounds obtained in this manner are subject to at least one order of magnitude in uncertainty, e.g. the upper bound is probably somewhere between $10^8$ and $10^7$ GeV~\cite{Chang:2018rso}.

\subsubsection{Axion helioscopes}

Helioscopes, such as the current one CAST~\cite{Anastassopoulos:2017ftl} and the future IAXO~\cite{Vogel:2013bta}, search for axions produced in the Sun~\cite{Sikivie:1983ip}.
Roughly speaking, they give $g_{a \gamma \gamma} \lesssim 10^{-10} / $ GeV and work for masses $m_a \lesssim eV$.
The basic principle used in these experiments is axion-photon conversion.  When two states mix, they can oscillate into each other, a phenomenon that is very familiar to physicists (e.g. particle physics see it in neutrino oscillations, whereas experimentalists see it in Rabi flopping).

Let's take a Hamiltonian with two states of energy $E_1$ and $E_2$ with a mixing term $\epsilon$.  It is a fun homework problem to work out that the mixing angle and eigenvalues obey
\bea
\frac{1}{2} \tan 2 \theta = \frac{\epsilon}{E_1-E_2} , \qquad E'_{1,2} = \frac{E_1 +E_2}{2} \pm \frac{E_1 - E_2}{2 \cos 2 \theta} .
\eea
Taking a state to be purely state 1 at some position, after a distance L, the probability of conversion into state 2 is
\bea
P(1 \rightarrow 2) = 2 \theta^2 \sin^2 \left ( \frac{L (E_1 - E_2)}{2} \right) .
\eea

In the presence of a large magnetic field, the $g_{a \gamma \gamma} a E \cdot B$ coupling in the Lagrangian causes mixing between the axion and photon.  For the axion,
\bea
\epsilon \sim g_{a \gamma \gamma} B , \qquad E_1 \sim \omega , \qquad E_2 \sim \sqrt{\omega^2 - m_a^2} .
\eea
Thus the axion photon conversion scales as
\bea
P \sim \left ( \frac{g_{a \gamma \gamma} B \omega}{m_a^2} \right )^2 \sin^2 \left ( \frac{L m_a^2}{\omega} \right) .
\eea
CAST utilizes a long region of space filled with a large magnetic field and searches for an axion produced in the sun that subsequently converts into a photon inside of CAST. 

\subsubsection{Light shining through walls}

A fun experiment called light shining through walls gives the constraint $g_{a \gamma \gamma} \lesssim 10^{-7} /$ GeV whenever $m_a \lesssim 10^{-3}$ eV~\cite{Ballou:2015cka}.  The basic idea is to shoot a laser at a wall through a region with a large magnetic field.  Behind the wall, there is another region with a large magnetic field followed by a photon detector.  When the laser approaches the wall, it can convert into an axion in the magnetic field.  The axion goes through the wall, whereas the photon hits it and is blocked.  After going through the wall, the axion converts back into a photon which can be measured.

\subsubsection{Polarization}

Only the photon mode that is transverse to a B-field can mix with the axion.  Thus, if a laser travels through a region with a large magnetic field, a birefringent effect arising from this mixing will mess with the polarization of the photon in the transverse-to-B-field direction.  Polarization-based experiments search for the change in polarization that results from this birefringent effect~\cite{DellaValle:2015xxa}.

\subsubsection{Black hole superradiance}

Black hole superradiance utilizes the interesting physics that happens around a spinning black hole~\cite{Arvanitaki:2009fg,Arvanitaki:2010sy,Arvanitaki:2014wva}, which has a region of space near to the horizon called the ergosphere where energies can be negative.  There is a famous process called the Penrose process by which an object is thrown into the ergosphere.  Inside the ergosphere, it imparts negative energy to some particle or part of the object, which is then thrown into the black hole while the rest of the object escapes.  The object leaves with more energy than it had when coming in, which extracts energy from the black hole.  Superradiance is a similar effect where energy is transferred from a black hole into an axion cloud surrounding it.  This process spins down the black hole so that high-spin black holes can be used to constrain the existence of axions.

\subsubsection{Fifth force experiments}

If $\overline \theta \ne 0$, then the axion has a coupling $\overline \theta  \psi \overline \psi a m_\psi/f$, which mediates a new Yukawa $\overline \theta^2/r^2$ force.  If the axion has spin couplings, $\overline \psi \gamma^\mu \gamma^5 \psi \partial_\mu a/f$, then it will mediate a $1/f^2 r^4$ force between spins.  If it has both of these couplings, then there is also a $\overline \theta/f r^3$ force between the two objects.  Experiments such as ARIADNE~\cite{Arvanitaki:2014dfa} are designed to look for this new $1/r^3$ force.

\subsubsection{Neutron star mergers at LIGO}

If the QCD axion is lighter than expected, then finite-density corrections to its potential can result in $\overline \theta = 
\pi$ being the minimum inside of a neutron star~\cite{Hook:2017psm}.  As a result, an axion field profile will 
surround a neutron star as $\overline \theta$ transitions from $\pi$ inside of the NS to $0$ outside.  Whenever two objects sourcing a 
field approach each other, there is a force between them.  Surprisingly, this new axion-mediated force can be as strong as gravity and either repulsive or attractive.  A new force of this type between neutron stars could be probed by LIGO~\cite{Huang:2018pbu}.

\subsubsection{Value of $\overline \theta$ in the Sun}

If $f_a \lesssim 10^{15-16}$ GeV and $m_a \lesssim 10^{-12}$ eV, 
finite-density corrections would cause $\overline \theta = \pi$ to be a minimum inside of the Sun~\cite{Hook:2017psm}.
Experiments have measured various nuclear properties related to the neutrinos and light coming from the Sun.  Hence we know that $\overline \theta \ne \pi$ in the Sun, and this can be used to place constraints on axions.

\subsection{Axion DM searches} \label{Sec: DMaxion}

In this subsection, I will discuss searches for axions/ALPs that rely on them being dark matter.  
I will only describe a subset of the numerous experimental probes available to us.

\subsubsection{How to treat axion/ALP DM}

The first question in axion DM searches is how to treat the axion.  Since we will only consider very light axions with $m_a < 0.1$ eV, the number of axions per Compton wavelength is large.  Hence we will treat the axions as a classical scalar field.

As the derivation is a bit tedious, we simply summarize the results :  The axion
acts like a classical field with $a(t) \approx \frac{2 \rho}{m^2} \cos m t$ for a coherence time $\tau \sim \frac{4 \pi}{m v^2}$.  After this time, the axion acquires a random phase and amplitude and is no longer acts as a cosine with the same phase for all time.
Thus, what one does conceptually is to divide time into slices of size $\tau$ and ``glue" them together and are combined in quadrature as these separate pieces are not phase coherent.

In this section, because we will use creation and annihilation operators, we will call our scalar field $\phi$ instead of $a$.
Phase experiments are sensitive to the value of $\langle \phi(t) \rangle$  while power experiments are sensitive to the value of $\langle \phi(t)^2 \rangle$.  To obtain these expectation values, we model dark matter as particles in a box and take the volume of the box to infinity.
We assume that the axions have some energy distribution given by dark matter simulations, typically isothermal, and ask how one calculates $\langle \phi(t) \rangle$ and $\langle \phi(t)^2 \rangle$.

At finite volume, the Hamiltonian is
\bea
H = \sum_n \omega_n a^\dagger_n a_n .
\eea
%whereas in infinite volume, we have
%\bea
%H = \int \frac{d^3 p}{(2 \pi)^3} \omega a^\dagger a
%\eea
%We have the mapping
%\bea
%\int d^3 p = \sum_n (\frac{2 \pi}{l})^3 \qquad a_n = \frac{a}{l^{3/2}}
%\eea
%when going between the two.
We decompose $\phi$ in terms of the creation and annihilation operators as
\bea
%\phi(x,t) &=& \int \frac{d^3 p}{(2 \pi)^3} \frac{1}{\sqrt{2 \omega}} ( a_p e^{i p \cdot x} + a_p^\dagger e^{-i p \cdot x} ) \\
% &=& \sum \frac{1}{l^{3/2}} \frac{1}{\sqrt{2 \omega_n}} (  a_n e^{i p \cdot x} + a_n^\dagger e^{-i p \cdot x}  )
\phi(x,t) = \sum \frac{1}{l^{3/2}} \frac{1}{\sqrt{2 \omega_n}} (  a_n e^{i p \cdot x} + a_n^\dagger e^{-i p \cdot x}  ) .
\eea

As mentioned before, people typically assume that the axion is a classical field.  A classical field is a superposition of a large number of particles and is an eigenstate of the creation and annihilation operators in the large $N_n$ limit.  Consider the state with an average number of particles $N_n$ with energy $\omega_n$,
\bea
\mid N \rangle = \alpha e^{\sqrt{N_n} a^\dagger_n } \mid 0 \rangle ,
\eea
where $\alpha$ is an unimportant normalization factor :
\bea
\langle N \mid H \mid N \rangle = \langle N \mid \sum \omega_n a_n^\dagger a_n \mid N \rangle = \sum \omega_n N_n .
\eea
Thus $N_n$ is the average number of particles in the $n$ state.  Because we are in a classical state, $N_n \gg 1$ so we will treat $a$ and $a^\dagger$ as if they commute.

Let us first express $N_n$ in terms of the distribution of particles in phase space.
Because particles come in waves, the energy at a fixed position is not constant in time. Let us define
\bea
\langle H \rangle_T = \frac{1}{T} \int dt H(t)  = \int d^3v f(v) \omega_v \overline n
\eea
Where $f(v)$ is the assumed classically derived distribution of axions.  It can now be easily shown that
\bea
\langle H \rangle_T &=& \sum \frac{1}{l^3} N_n \omega_n , \qquad N_n = ( \frac{2 \pi}{m} )^3 f(v_n) \overline n .
\eea 
This is exactly what we expect, as it's just the total energy divided by the volume.

We now expand around a random position $x_0$ and a time $t_0$.  This means that all of the relative phases between the sums will be basically random.  For simplicity, we assume that $f$ is isotropic so that it is independent of the angles :
\bea
\langle \phi(t) \rangle = \sum_{i_v,i_\theta,i_\phi} ( \frac{2 \pi}{m l} )^{3/2} \sqrt{\frac{2 \overline n f_{i_v}}{\omega_{i_v}}} \cos \left( \omega_{i_v} t + \phi_{i_v} + \phi_{i_\theta} + \phi_{i_\phi} \right ) .
\eea
The velocity distribution $f_{i_v}$ has a characteristic scale $v_0$.  In particular, for changes in velocity $\Delta v \ll v_0$, we can treat $f_{i_v}$ and the cosine piece as approximately constant.  
Because of this approximation, we go from being able to trust the time evolution for all time, to being able to trust it only for a time $2 \pi/(m v_0 \Delta v)$.
We thus separate the sum into regions where $f_{i_v}$ is constant,
\bea
i_v = a_v \frac{\Delta v l m}{2 \pi} + b_v .
\eea
We will approximate $f_{i_v}$ and $\omega_{i_v}$ as functions of $a_v$ only and the energy in the denominator as the mass as it varies more slowly than $f$, so that
\bea
\langle \phi(t) \rangle \approx \sum_{a_v}  ( \frac{2 \pi}{m l} )^{3/2} \sqrt{\frac{2 \overline n f_{a_v}}{m}} \sum_{b_v}^{\frac{\Delta v l m}{2 \pi}} \sum_{i_\theta,i_\phi} Re \left ( e^{i (\omega_{a_v} t + \phi_{{a,b}_v} + \phi_{{a,b}_\theta} + \phi_{{a,b}_\phi}) } \right ) .
\eea
The latter part of this expression can be simplified by noting that the sums over $b,i_\theta,i_\phi$ are just a random walk, wherethe number of steps is the number of phases in the volume
\bea
N = 4 \pi i_r^2 \frac{\Delta v l m}{2 \pi} = 4 \pi v^2 \Delta v ( \frac{l m}{2 \pi} )^3 .
\eea
Summing over this many random phases gives us
\bea
\langle \phi(t) \rangle  = \frac{\sqrt{\overline \rho}}{m} \sum_{a_v}  \sqrt{f_{a_v} 4 \pi v^2 \Delta v} \alpha_{a_v} \cos \left ( \omega_{a_v} t + \phi_{a_v} \right ) ,
\eea
where the random numbers $\alpha_{a_v}$ are taken from the Rayleigh distribution, 
\bea
P(\alpha_{a_v}) = \alpha_{a_v} e^{-\alpha_{a_v}^2/2} .
\eea
The expectation value of $\langle \phi(t)^2 \rangle$ is much easier to calculate and we leave it as an exercise for the reader.

We are finally in the position to say how experiments should treat the expectation values
\bea
\langle \phi(t) \rangle = \frac{\sqrt{\overline \rho}}{m} \sum_{i_v}  \sqrt{f_v 4 \pi v^2 \Delta v} \alpha_r \cos \left ( \omega_r t + \phi_r \right ) , \\
\langle \phi(t)^2 \rangle = \frac{\overline \rho}{m^2} \sum_{i_v}  f_v 4 \pi v^2 \Delta v \cos^2 \left ( \omega_v t + \phi_v \right ) .
\eea
These sums converge independently of the value of $\Delta v$ as long as $\Delta v$ is small.  In practice, it is simplest to take $\Delta v$ to be the resolution of the experiment.
We can now simulate the above field values and check that they are well approximated by a cosine during a coherence time $\tau \sim \frac{4 \pi}{m v^2}$.  For times longer than a coherence time, $\phi(t)$ is no longer well approximated by a cosine leading to the behavior described earlier in this subsection.

\subsubsection{Astrophysical probes}

There are a slew of astrophysical probes of the axion and ALP DM, which look for DM-photon conversions in galactic magnetic fields, around NS, and at CMB to name a few.  As you'll hear more about astrophysics and DM in your other lectures, I won't go into further detail here.

\subsubsection{Haloscopes}

Haloscopes, the prototypical example being ADMX~\cite{Asztalos:2009yp}, search for evidence of DM converting into photons in
a magnetic field~\cite{Sikivie:1983ip}.  If we are looking for conversions in a cavity, we have the cavity modes $- \nabla \psi_m = \omega_m^2 \psi_m$ and orthonormality conditions $\int \psi_n \psi_m dV = V \delta_{nm}$.  Maxwell's equations for the modes are
\bea
\frac{\partial^2 B}{\partial t^2} + \nabla^2 B = - g_{a \gamma \gamma} B_\text{ext} \ddot a.
\eea
In terms of the cavity modes, we have
\bea 
\label{Eq: experiment}
(\omega^2 - \omega_n^2 + i \omega_n \Gamma_n) B_n = g_{a \gamma \gamma} B_\text{ext} \omega_n^2 a_0 \int \frac{dV}{V} \psi_n \cdot B_\text{ext} \cos \left ( m_a t + k_a x \right ) .
\eea
We see from the last term that we want the cavity mode to overlap as much as possible with the axion wavelength.  This generally means that $L \sim 1/m_a$, so that the cavity is of order the size of the axion, and that conversion is only efficient for the fundamental mode.

Axion-photon conversions are all about matching the dispersion relationships in space, so the size of the experiment is about 
the size of the axion, and in time, the energy of the mode is about the energy of the axion.  Hence ADMX looks for the 
energy deposited by the axion into the cavity.

\subsubsection{ABRACADABRA}

Examining Maxwell's equations
\bea
\nabla \times B = \frac{\partial E}{\partial t} + J + \frac{\dot a}{f} B ,
\eea
we notice that the axion acts like an effective current along B-fields.  Consider now a solenoid with current
running through it.  In pure E+M, there is a B-field along the solenoid and 0 B-field outside.  In the presence
of an axion, however, this B-field inside acts like a small current generating a non-zero B-field outside.  ABRACADABRA~\cite{Kahn:2016aff,Ouellet:2018beu} 
searches for this non-zero B field outside of a solenoid bent into a toroid.  

\subsubsection{CASPER}

CASPER~\cite{Budker:2013hfa} uses NMR to hunt for the axion.  In the presence of the axion,
the neutron has a time-dependent eDM.  Consider a spin-polarized block of material
that is roughly polarized in the $z$ direction.  Apply a B-field in the $z$ direction and an E-field in the $y$ direction.
The spin will precess around the $z$ axis with a frequency $\omega_B = g B$, where $g$ is the gyromagnetic ratio.
In the absence of the axion, the slightly misaligned spins precess around the $z$ axis and stay only slightly misaligned.

Now, if the axion is present, there is also an eDM aligned with the spin that causes precession around the $y$ axis.  Because the axion oscillates in time, this precession never performs a full oscillation and only rocks back and forth at a frequency $m_a$.  I encourage the reader to take pencil and paper to sketch this precession around the $z$ axis and
small wobbles about the $y$ axis.  Using geometry, it can be seen that the spin becomes more and more misaligned with the $z$ axis as time progresses if $m_a \sim \omega_B$.  Basically, each effect oscillates as $\cos \omega t$ so that each individually averages to zero as time progresses.  However, the combined effect $\cos \omega_B t \cos m_a t$ will grow with time if $m_a \sim \omega_B$. CASPER is designed to look for this effect.

\subsubsection{Dielectric haloscopes}

MADMAX~\cite{TheMADMAXWorkingGroup:2016hpc} and its cousin~\cite{Baryakhtar:2018doz} are a set of proposed
experiments that use dielectrics.  The idea is to notice that in Eq.~\ref{Eq: experiment}, the RHS involves an integral
over the wavefunctions of the cavity mode and the axion mode.  Normally cavity wavefunctions scale as
 $\sin ( n \pi x/L )$.  If the experiment has a size much smaller than the de Broglie wavelength of the axion, so that we can ignore any spatial dependence of the axion, the higher modes start integrating to zero.  Dielectric haloscopes use dielectrics to modify these wavefunctions so that the wavefunctions are no longer pure cosines and sines and $\int \psi_n dx$ does not fall off as $n$ increases.
% 
% there is a long wavelength mode when the wavefunction $ > 0$, and then when the wavefunction $< 0$, changes the wavelength to make it very short.  
Since the integral of this wavefunction over many periods does not vanish, these higher-order modes can be used for axion-photon conversion due to the fact that the integrals are enhanced by the length of the material $L$.  These experiments look for photons from axion-photon conversion that are emitted from series of stacked dielectrics.

\subsubsection{Interferometers}

Again, by looking at Maxwell's equations, we see that the dispersion relation for circularly polarized light is
\bea
\omega^2 = k^2 \pm \frac{k \dot a}{f}.
\eea
Left and right circularly polarized light travel at different phase velocities.  Thus the natural thing to do is to build an interferometer to look for the different phase velocities~\cite{DeRocco:2018jwe,Liu:2018icu}.  This is exactly same way that we look for gravity waves
except that gravity-wave interferometers are insensitive to the polarization of the light.

\section{Conclusion}

Hopefully these sets of lectures have been a useful introduction to the exciting field of the Strong CP problem and axions.  Their target audience is graduate students who are interested in doing research in this area.  Of course, many topics and references have been left out due to lack of time and due to my lack of energy.  This is a bustling field of research which has recently picked up a lot of momentum.  On the theory side, model builders are expanding upon the basic minimal parity and axion solutions.  On the experimental side, there have recently been many new proposals for how to search for axion dark matter.  These ideas are very exciting and prototypes of many of these proposals are being built.

\section*{Acknowledgments}

I thank everyone involved in TASI 2018 for making it such a great experience.  I thank Yvonne Kung and Gustavo Marques Tavares for carefully reviewing the manuscript and fixing the many typos and errors within it.

\bibliography{biblio}

\end{document}